\documentclass[10pt,twocolumn]{article}
\usepackage[margin=0.75in]{geometry}
\usepackage{color}
\usepackage[normalem]{ulem}
\usepackage{amsmath,amsthm, amssymb}
\usepackage{tcolorbox}

\usepackage{graphicx}
\usepackage[hyperfootnotes=false]{hyperref}
\usepackage{xurl}
\usepackage{authblk}

\usepackage{subfig}
\usepackage{multirow}
\usepackage{array}
\usepackage{bold-extra}
\usepackage{bm}
\usepackage[numbers,square]{natbib}
\usepackage{algorithm}
\usepackage{algpseudocode}
\usepackage{abstract}

\begin{document}




\title{Estimating the cascading global impacts of gas disruptions in Qatar}
\date{}

\author[1,2]{Ritwick Mishra}
\author[1]{Diksha Gupta}
\author[1,3]{Achla Marathe}
\author[4]{Krista Danielle Yu}
\author[1]{Aaron Schroeder}
\author[1]{Samarth Swarup}
\author[1]{Brian Klahn}
\author[5]{Phil Potter}
\author[1,2]{Madhav Marathe}
\author[1,2,$\ast$]{Anil Vullikanti}

\affil[1]{Biocomplexity Institute, University of Virginia}
\affil[2]{Department of Computer Science, University of Virginia}
\affil[3]{Department of Public Health Sciences, University of Virginia}
\affil[4]{Department of Economics, De La Salle University}
\affil[5]{National Security Data and Policy Institute, University of Virginia}



{



\twocolumn[
  \maketitle 
  \vspace{-4em}
  \begin{center} 
      \footnotesize{$^\ast$\textit{Correspondence email}: \texttt{vsakumar@virginia.edu}}
\end{center}
  \begin{onecolabstract}
This study examines the global impacts of a localized disruption in
Qatar’s gas sector using a multi-regional input–output framework and
scenario-based analysis. 
While the direct impacts of this disruption on importing countries are clear, indirect and cascading impacts are not well understood.
We use a Multiregional input-output (MRIO) model to assess the impact of this disruption and to determine whether trade reallocations and increased production can mitigate its effects.
Our analysis shows that this disruption leads to significant gas
supply losses in Asia and Europe, with the largest aggregate impacts
observed in India, China, and South Korea. 
Allowing for trade reallocation partially
mitigates these losses.
Further expansion of production capacity among major gas-producing countries improves supply conditions and leads to broader output gains; however, these benefits remain concentrated in a few large economies. Even significant increases in production among top producers offer limited relief to economies such as India and Pakistan.
Overall, the results highlight the uneven distribution of both
vulnerabilities and recovery potential within global supply
chains. While adaptive mechanisms such as trade reallocation and
production expansion can alleviate the effects of supply shocks, their
effectiveness is limited and heterogeneous. The findings underscore
the importance of network structure in shaping shock propagation and
resilience, offering insights for
managing systemic risks in an interconnected global economy.
  \end{onecolabstract}
  \vspace{2em}
]
\newcommand{\anil}[1]{\textcolor{blue}{#1}}
\newcommand{\samarth}[1]{\textcolor{orange}{(SS: #1})}
\newcommand{\madhav}[1]{\textcolor{green}{#1}}
\newcommand{\achla}[1]{\textcolor{red}{#1}}
\newcommand{\krista}[1]{\textcolor{magenta}{#1}}
\newcommand{\rmcomment}[1]{{\noindent \textcolor{teal} {\textbf{RM}: {#1}}}}

\newcommand{\dgcomment}[1]{{\noindent \textcolor{purple} {\textbf{Diksha}: {#1}}}}



\date{}

\section{Introduction and Background}



Qatar is one of the world's largest exporters of liquefied natural gas (LNG), with production centered at Ras Laffan Industrial City, the largest LNG complex in the world. Accounting for 20\% of globally traded LNG, Qatar is a critical node in the energy security of major importing economies across Asia (China, India, Japan, South Korea) and Europe, making international markets highly sensitive to any disruption in its output.

Most Qatari exports transit through the Strait of Hormuz, a narrow and geopolitically exposed corridor that adds considerable systemic risk to global energy supply. In early March 2026, Iranian strikes on key LNG infrastructure at Ras Laffan and Mesaieed took approximately 17\% of Qatar's LNG export capacity offline, equivalent to roughly 13 million tons per year, with recovery timelines potentially extending over several years~\cite{iea2024weo}.

The disruption to Qatar's gas sector is reverberating across the world, with Asia and Europe bearing the heaviest burden. In India and Pakistan, where LNG and liquefied petroleum gas (LPG) imports are heavily sourced from Qatar and routed through the Strait of Hormuz, supply tightening has already disrupted LPG distribution, leaving millions unable to cook and forcing the closure of small businesses and restaurants~\cite{anand2026}. The Philippines has declared a national emergency in response to rapidly rising fuel prices and introduced shortened workweeks, while Pakistan has closed schools to conserve energy~\cite{nyt-gas-runs-out}.

The reported effects extend to countries with no direct trade relationships with Qatar.
For example, even though the US is a major producer and net exporter of gas, a disruption in Qatar’s gas sector results in higher prices for American households. When Qatari supply falls, global gas prices rise, causing US gas to be diverted to higher-paying export markets.

Despite the scale of these disruptions, the cascading second- and third-order effects on the global economy remain poorly understood. Beyond Qatar, major LNG producers include Australia, Norway, Russia, Canada, and Algeria, and a key open question is whether these countries can fill supply gaps through export reallocation or increased production. Addressing this requires modeling tools capable of capturing the complex interdependencies that allow shocks to propagate across sectors and countries.

Multiregional input-output (MRIO) models are widely used for this purpose, as they explicitly represent how production linkages transmit shocks through global supply networks~\cite{galbuseraandgiannopoulos2018}. Firm-level and network-based studies confirm that disruptions can generate substantial indirect effects, particularly in supply chains characterized by limited substitutability and dense interconnections~\cite{inoueandtodo2019,zhangetal2025}. Extensions of MRIO frameworks to energy systems have further shown that interregional trade can partially offset demand deficits following production disruptions, though the degree of mitigation depends critically on network structure and capacity constraints~\cite{heetal2019}.
 
These approaches have important limitations. Conventional MRIO models assume fixed production structures, precluding trade reallocation across alternative suppliers under binding constraints, and can yield economically infeasible outcomes when supply and demand shocks occur simultaneously~\cite{pichler2022simultaneous}. Koks and Thissen~\cite{koksandthissen2016} address some of these shortcomings by combining multiregional supply-use structures with cost-minimizing optimization, incorporating supply constraints, production inefficiencies, and substitution across suppliers. However, in their framework reallocation is driven by optimization objectives rather than grounded in existing trade relationships and capacity constraints. As a result, neither conventional MRIO models nor optimization-based extensions fully capture how production and trade adjust when reallocation is shaped by real-world structural constraints.

In this study, we develop a global MRIO optimization model that incorporates disruption, reallocation, and capacity expansion using the GTAP 12 database. 
We use it to examine how shocks to Qatar's gas sector (\texttt{QAT-GAS}) propagate through international supply chains.

We construct three hypothetical disruption scenarios: varying levels of gas disruption, partial reallocation by six largest gas-producing countries (\texttt{TOP-GAS}) outside the Gulf region, and increased gas production by these same countries.
Together, these scenarios allow us to unravel the relative
importance of supply shocks, trade reallocation, and production
responses. 
Our modeling framework captures both direct effects (reduced gas supply to
importing countries) and indirect effects (downstream impacts on
industries that rely on gas as an input).
A reduction in gas supply not only affects the energy
sector; it cascades through the economy, leading to broader output and final demand
losses.  We consider a global networked trade flow model where we model the industries and final demand region of each country as nodes and flow of goods between them as edges with capacities.
Using this networked model of trade flow, the  effects of a localized gas disruption can be traced across sectors and
countries, showing how a single, highly connected node, in our case, Qatar’s gas
sector, can generate ripple effects reaching nearly every region in the world.
Furthermore,
by comparing outcomes across scenarios, we can assess the extent to
which global supply chains exhibit resilience or vulnerability, and
identify key regions and sectors that are most exposed to such
shocks. This approach provides a systematic basis for understanding
the broader economic implications of energy supply disruptions and the
potential effectiveness of alternative adjustment mechanisms.

\section{Results}

\begin{table*}[tb]
\centering
\begin{tabular}{|l|c|c|c|}
\hline
\textbf{Component} & \textbf{Scenario 1} & \textbf{Scenario 2} & \textbf{Scenario 3} \\
\hline

\multicolumn{4}{|c|}{\textit{Disruption: Reduction in Flow Capacity (\texttt{QAT-GAS})}} \\
\hline
Edges affected & \texttt{QAT-GAS $\rightarrow$ ALL} & \texttt{QAT-GAS $\rightarrow$ ALL} & \texttt{QAT-GAS $\rightarrow$ ALL} \\
Magnitude ($\delta$) & $\{{\bf 0.3}, 0.5, 0.7\}$ & ${\bf 0.3}$ & ${\bf 0.3}$ \\
\hline

\multicolumn{4}{|c|}{\textit{Adaptation: Increase in Flow Capacity}} \\
\hline
Edges affected & -- & \texttt{TOP-GAS $\rightarrow$ ALL} & \texttt{GLOBAL} \\
Magnitude ($\alpha$) & -- & $\{0.1, 0.2, {\bf 0.3}\}$ & $\{{\bf 0.05}, 0.1, 0.2\}$ \\
\hline

\multicolumn{4}{|c|}{\textit{Adaptation: Increase in Production Capacity through primary inputs}} \\
\hline
Firms affected & -- & -- & \texttt{TOP-GAS} \\
Magnitude ($\beta$) & -- & -- & $\{{\bf 0.05}, 0.1, 0.2\}$ \\
\hline

\end{tabular}
\caption{Summary of disruption and adaptation scenarios. We consider three disruption scenarios: (1) All trade
linkages (edges) originating from Qatar’s gas sector are reduced
uniformly by $\delta = 30\%, 50\%, 70\%$, with no mitigation; (2) To mitigate the effect of disruption, the six largest gas-producing countries can have higher flow capacities by $\alpha$ = 10\%,
20\%, 30\% but production remains the same; (3) These same producers
can increase their primary inputs by $\beta=5\%, 10\%, 20\%$ in the gas
sector to help increase production. To accommodate the additional production, enabled by
the increase in primary resources, global trade edge capacities ($\alpha$)
are increased proportionally i.e. $\alpha=\beta$. The settings in bold correspond to the scenarios used for the heatmaps.} 
\label{tab:scenario}
\end{table*}


\begin{figure*}[t]
    \centering
    \includegraphics[width=0.96\linewidth]{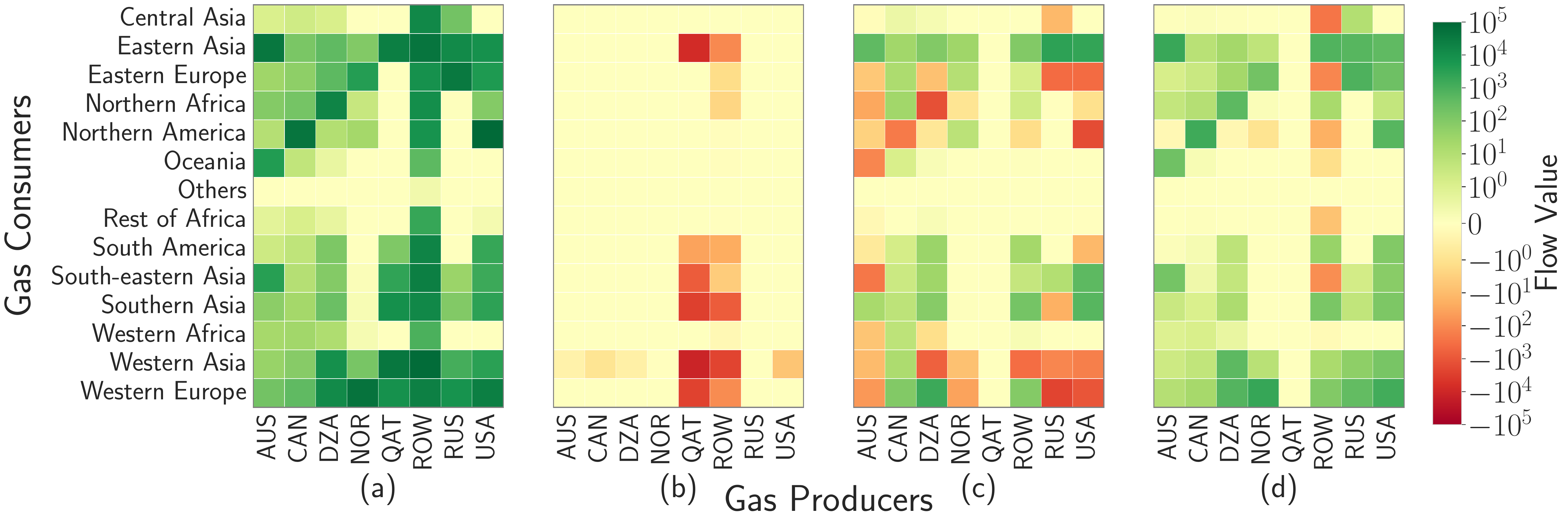}
    
    
    \caption{Heatmap of gas flows from producer countries (x-axis) to consumer (y-axis) regions. (a) Original gas flows in the MRIO table. (b) Scenario 1 $(\delta=0.3)$: Change in gas flows with respect to original values ($\delta=0$) (c) Scenario 2 $(\delta=0.3, \alpha=0.3)$: Change in gas flows with respect to Scenario 1 $(\delta=0.3)$. (d) Scenario 3 $(\delta=0.3, \alpha=0.05, \beta=0.05)$: Change in gas flows with respect to Scenario 1 $(\delta=0.3)$. All figures are in mUSD.
    }
    \label{fig:gas_flow_mat_htmp}
\end{figure*}

We consider three  disruption scenarios as described in Table \ref{tab:scenario}:
(1) All trade linkages (edges)
originating from Qatar’s gas sector have their capacities reduced uniformly by 30\%, 50\%,
and 70\%, representing varying levels of supply contraction. This scenario captures the shock to global gas supply and its cascading effects on dependent sectors and countries, with no mitigation;
(2) The six largest gas-producing countries outside the
Gulf region partially reallocate their exports to
global markets, with adjustments of 10\%, 20\%, and 30\% of the flow, reflecting
trade diversion and substitution effects under constrained supply conditions;
(3) These
same producers increase their primary inputs by 5\%, 10\%, and 20\% in the gas sector. To
accommodate the additional production, enabled by the increase in primary resources, global trade edge capacities
are increased proportionally, enabling the redistribution of supply
across the network.

 We develop a linear programming (LP) approach to find solutions to MRIO models for these scenarios using the GTAP 12 database, which covers 163 countries and 65 sectors (see Section \ref{sec:method} for details).
This is used for the industry-to-industry flow linkages within each country and between countries to build our networked trade model. The sectors of each country (e.g., US-Gas, China-electronics) form the nodes, and the flow linkages form the directed edges, whose capacities are originally set to the MRIO values in USD. Additionally, there is one `sink' node per country to represent the final demand. 
We analyze properties of a solution to the LPs for these scenarios.
While these are feasible solutions, we note that solutions need not be unique, and some country specific properties might vary for other solutions.

Throughout the text, we denote the six Gulf countries of Saudi Arabia, Qatar, UAE, Kuwait, Bahrain, and Oman together as \texttt{GULF}. Also, we denote the set of six top gas-producing sectors by gross output, excluding \texttt{GULF}, as \texttt{TOP-GAS}, namely, those of US, Russia, Norway, Canada, Australia and Algeria. 


\subsection{Scenario 1: Disruption without any mitigation}

We model the disruption by proportionally reducing the flow capacities of edges from \texttt{QAT-GAS}  to all 
other nodes, controlling the disruption percentage with a parameter $\delta$. In this baseline scenario, all 
other flow capacities remain unchanged, which implies that the flows cannot exceed their original values to mitigate the demand 
loss. In Figure~\ref{fig:gas_flow_mat_htmp}(a), we show the original gas flows from major gas producers to 
different regions of the world as a heatmap, illustrating the dependency structure on gas. These gas flows are used either as intermediate inputs (i.e., as input to other firms) or as final demand by the households. 
Figure~\ref{fig:gas_flow_mat_htmp}(b) reports the loss of gas flows due to the disruption (i.e., the original 
gas flows minus the flows under disruption).
The impacted regions are largely those that were 
directly dependent on Qatar gas. 
While the impact is global, it is most concentrated in Asia, where
several regions experience gas supply shortfalls of up to $10$
billion USD\footnote{All USD values reported are in 2023 USD.}.
Western Europe also incurs losses on the order of
billions. Beyond Qatar, gas flows from the rest of the world
(ROW), which includes other Gulf producers, are affected to a lesser extent,
indicating indirect dependence on Qatari gas as an input into their own production. This secondary disruption further amplifies losses,
particularly across Asia and Europe. 

Next, we examine the cascading effects of these gas-flow disruptions on the overall economies of countries around the world.
Figure~\ref{fig:heat_delta} shows the negative changes in final demand and gross output  (compared to the original values, i.e.,~Scenario 1~-~Original) aggregated at a country level and visualized as global heatmaps. Nearly all countries are affected to some degree, but regions in Asia and Europe are affected the most, aligning closely with the gas flow losses seen in Figure~\ref{fig:gas_flow_mat_htmp}(b). 
In terms of loss in final demand, the large Asian economies of India, China, Korea are heavily affected, 
followed by Italy and the US. This is directly connected to the loss of gas supply from Qatar, as seen in 
Figure~\ref{fig:gas_flow_mat_htmp}(b). However, the US is an exception here, as the direct flow of Qatar gas to the 
US is negligible. This demonstrates the impact of indirect channels of disruption, for example, the directly affected Asian sectors propagating production losses to US customers. Output losses are somewhat more contained but have 
overlapping geographic regions, with additional impacts observed in Taiwan. This suggests that gas from Qatar  is a critical input to industry in these regions, and that a disruption in this input can cascade through 
their highly integrated internal economies. Figure~\ref{fig:vary_delta} shows the aggregate losses for the top 5 
countries as the disruption parameter $\delta$ varies. As $\delta$ increases from $0.3$ to $0.5$, the losses 
nearly double for India and China, whereas losses in other countries rise more gradually. This difference 
reflects the varying dependence on gas imports and downstream processing capacity across economies.


\subsection{Scenario 2: Disruption combined with reallocation}

In this scenario, we assume that the outgoing edges 
from \texttt{TOP-GAS} (a set of top six gas-producing firms by gross output excluding \texttt{GULF}) can change their flow capacity by $\alpha \%$. That is, the flow on these outgoing edges can increase by at most $\alpha \%$ from its original value (i.e., at $\delta=0$), and decrease by at most $\alpha \%$ below its Scenario 
1 value\footnote{The limit on the decrease in flow is imposed to ensure that the new solution does not deviate 
too much from the solution obtained in Scenario 1.}.
Note that the disruption remains the same as in Scenario 1.
Further, for all edges (other than the outgoing edges from \texttt{TOP-GAS}), the flow capacity remains the same as before.


Note that in this scenario, we assume that these sectors do not increase their production as they are still limited by their inputs, but can only reallocate out-flows to their recipient firms. 
Our linear programming approach (Section \ref{sec:method}) finds the optimal re-allocation under these constraints.
Figure~\ref{fig:gas_flow_mat_htmp}(c) shows the change in distribution of gas flows to different regions (compared to Scenario 1, i.e., Scenario 2  - Scenario 1) when reallocation is allowed by \texttt{TOP-GAS}. 
We find significant reallocation of gas to the affected regions, with Asia and Western Europe being the major beneficiaries. Regions in red see a decline in flow and regions in green are gaining flow. While Eastern Asia sees an increase in gas flows from all the major gas producers, especially Russia and the US (also stated in articles~\cite{reuters_asia_us_lpg_2026, china_russia_energy}), Western Europe relies more on increased gas supplies from Algeria and Canada. The US redirects its gas flows mostly towards Asia instead of Europe. Notably, Western Asia benefits little from the reallocation, with Southern Asia also losing out to East Asia in attracting
larger gas flows. Australia also redirects its flows to Eastern Asia and away from the rest of the regions. 
Notably, these additional flows are coming at the cost of domestic supply in \texttt{TOP-GAS} regions, which can been seen in the declines to Oceania, North America, North Africa, Eastern Europe etc. In this scenario, the optimal strategy tends to partly redirect domestic supply towards critically unmet demand abroad. 
This suggests that when given additional flow capacities, gas suppliers prioritize systemically important sectors in the world, which are often located in East Asia and Western Europe, if the goal is to maximize the overall output and final demand in the world. 

When we assess the global economic losses in Figure~\ref{fig:heat_alpha}, we see that the increase in flow capacity leads to wider improvements in output and final demand in Asia and Southern Europe. 
Countries such as China, Korea, and Italy benefit from increased gas flows through higher economic output.
In \texttt{TOP-GAS} countries (e.g, US, Canada, Australia), redirecting gas away from domestic consumption toward foreign unmet demand (shown in Figure~\ref{fig:gas_flow_mat_htmp}(c)) yields a net positive effect: the aggregate demand and output of their nation-wide economies increases above the baseline values in Scenario 1 (Figure~\ref{fig:heat_alpha}). This demonstrates that prioritizing exports to ``bottleneck'' firms abroad over domestic supply can lead to aggregate gains in the domestic economies. This suggests that redirected flows propagate through global supply chains, ultimately generating increased economic demand and output in the exporting countries.
Figure~\ref{fig:vary_alpha} shows the loss in aggregate final demand and outputs for the top 5 countries as the 
reallocation parameter $\alpha$ varies, while keeping the disruption parameter $\delta$ fixed. We observe that 
the recovery due to this adaptation is not uniform. Countries like the US, Korea, and China reduce their losses 
by up to two orders of magnitude as $\alpha$ is doubled from $0.1$ to $0.2$. This contrasts with India, which 
sees much more modest improvement, while Pakistan shows no improvement despite the possibility of reallocation. 


\subsection{Scenario 3: Disruption mitigated through increased gas production}

In this scenario, we assume that \texttt{TOP-GAS} firms increase their primary input by $\beta\%$ to help 
fulfill the unmet demands due to the disruption. For any given firm, effective utilization of this additional 
capacity requires corresponding increases in both intermediate inputs and
outbound flow capacities. In the absence of sufficient downstream flow
capacity, firms may be unable to realize higher production levels
without generating excess or unused output.

Accordingly, we first augment the primary inputs of the
\texttt{TOP-GAS} firms by $\beta$\%. Even a localized increase in demand or production can
propagate through the network, inducing higher input requirements
across interconnected sectors and regions. This creates a non-trivial
allocation problem:  which specific edge capacities should be
expanded to support the intended production increase? To address this,
we adopt a simplified approach: we allow a uniform expansion of flow
capacities across all edges, while restricting the increase in primary
inputs to the selected \texttt{TOP-GAS} firms. This ensures that the
augmented production can be effectively distributed throughout the
network without binding capacity constraints.
Just like in Scenario 2, we control the reallocation by setting $\alpha=\beta$. Each edge can increase its flow by $\alpha$\% over its original value and can decrease by up to $\alpha$\% below its Scenario 1 values. 

Figure~\ref{fig:gas_flow_mat_htmp}(d) shows the change in distribution of gas flows from major producers to different regions as compared to the Scenario 1 values. With additional production capacity, all \texttt{TOP-GAS} countries improve their supply to affected regions as well as to their own economies. Invariably, domestic gas demand improves significantly in each producer country. The gains for Eastern Europe and Western Asia intensify as compared to Scenario 2 (in Figure~\ref{fig:gas_flow_mat_htmp}(c)). Differing from Scenario 2, there is increased Australian gas flow to Southeastern Asia, Russian gas to Eastern and Western Europe, and American gas to South America and Western Europe.
Western Asian gas supply is now much improved compared to Scenario 2, bolstered with additional flows from Algeria and the US. There is only a modest increase in flows to Southern Asia, even though the region faces large shortfalls in gas.

In Figure~\ref{fig:heat_beta}, we plot the global heatmap of changes in final demand flows and gross output per country. We measure these changes relative to Scenario 1 values.
The combination of production expansion and an increase in global flow capacity substantially mitigates
losses, particularly in large economies such as the US,
China, and India. At the same time, reallocation redistributes
impacts on final demand, with countries in the Americas experiencing
relatively greater losses compared to Scenario 1. 
Gains in gross output are observed
across much of the world, including heavily affected economies such as China, India, and Italy. This suggests that enabling reallocation over all edges can lead to higher intermediate demand relative to final demand, even though we give equal weights to both in our optimization objective.
\texttt{TOP-GAS} group, the benefits accrue
disproportionately to countries with strong integration into global gas-dependent supply chains. In particular, economies such as India, China, and South Korea capture a significant share of the gains,
alongside the \texttt{TOP-GAS} countries themselves.
Figure~\ref{fig:vary_beta} shows the sensitivity of the losses of the 5 most affected countries to the 
parameter $\beta$.  Similar to Scenario 2, the recovery is heterogeneous with countries like India and Pakistan 
improving only modestly with the increase in $\beta$, while others like China and Korea see substantial 
improvements.
Notably, China’s output losses decline sharply, from approximately
630 billion USD to nearly zero as $\beta$ increases from 0.05 to 0.1, underscoring the uneven distribution of resilience across economies.



\begin{figure}
    \centering

    \includegraphics[width=1.0\linewidth]{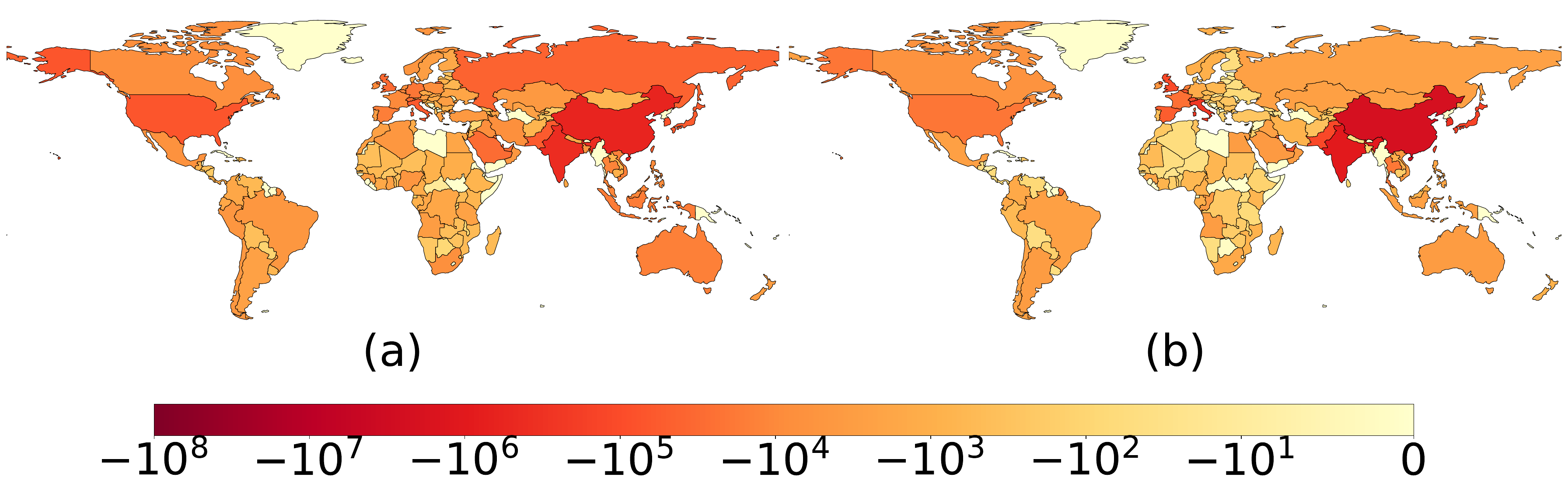}
    \includegraphics[width=1.0\linewidth]{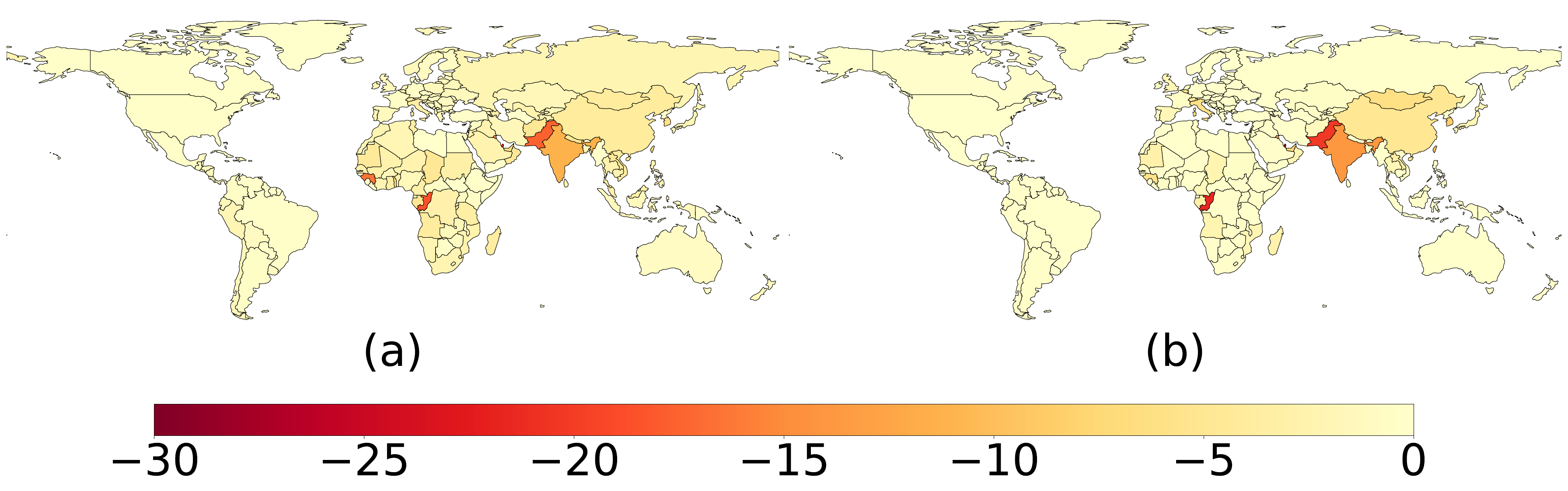}
    
    
   
    
    
    \caption{Scenario 1 $(\delta=0.3)$: Heatmap of changes in aggregate (a) final demand, (b) output with respect to original values ($\delta=0)$. Top plots are in mUSD. Bottom plots are in percentages of original values.}
    \label{fig:heat_delta}
\end{figure}

\begin{figure}
    \centering
    \subfloat[]{\includegraphics[width=0.38\linewidth]{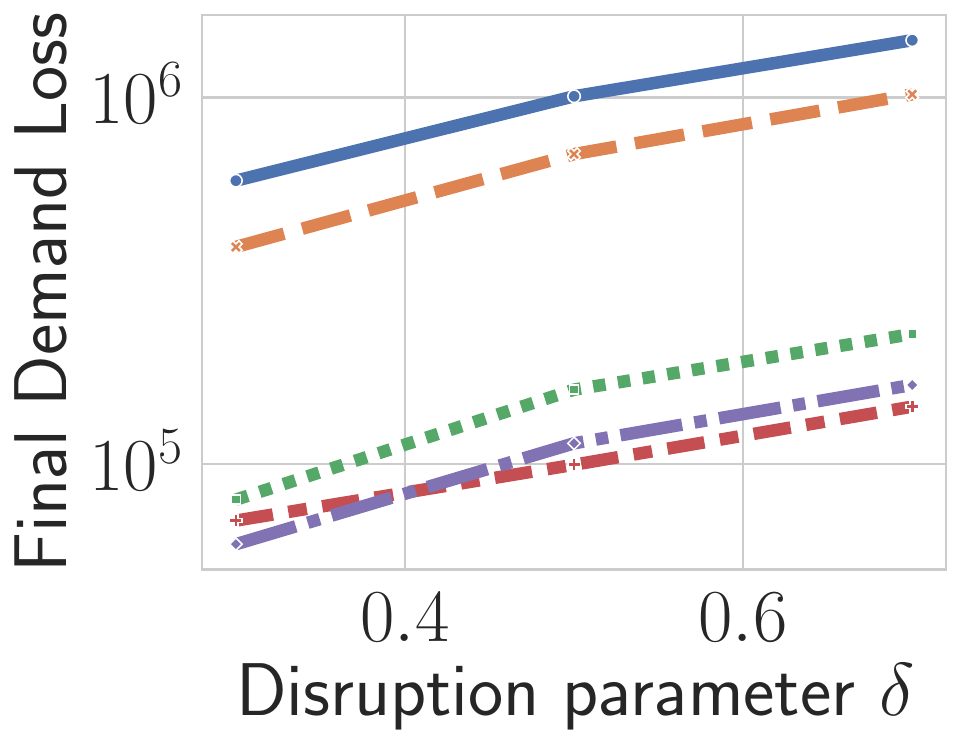}
    \includegraphics[width=0.1\linewidth]{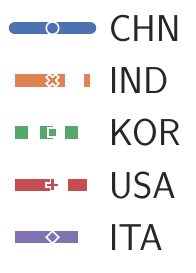}}
    \subfloat[]{\includegraphics[width=0.38\linewidth]{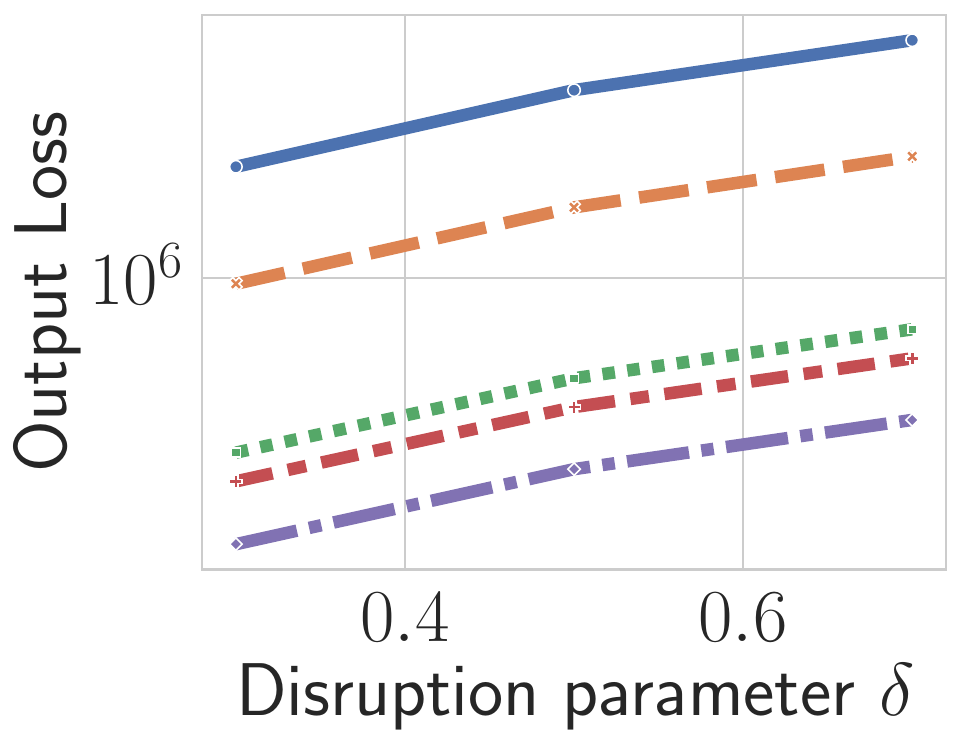}
    \includegraphics[width=0.1\linewidth]{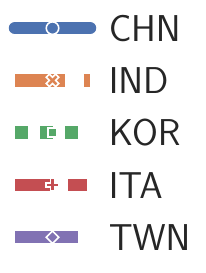}}
    \caption{Scenario 1 $(\delta=\text{var.})$: Top 5 most impacted countries (excl. \texttt{GULF}) in terms of loss in (a) final demand, (b) output, with respect to original values with varying disruption 
    parameter $\delta$. All figures are in mUSD.}
    \label{fig:vary_delta}
\end{figure}

\begin{figure}
    \centering
    \includegraphics[width=1.0\linewidth]{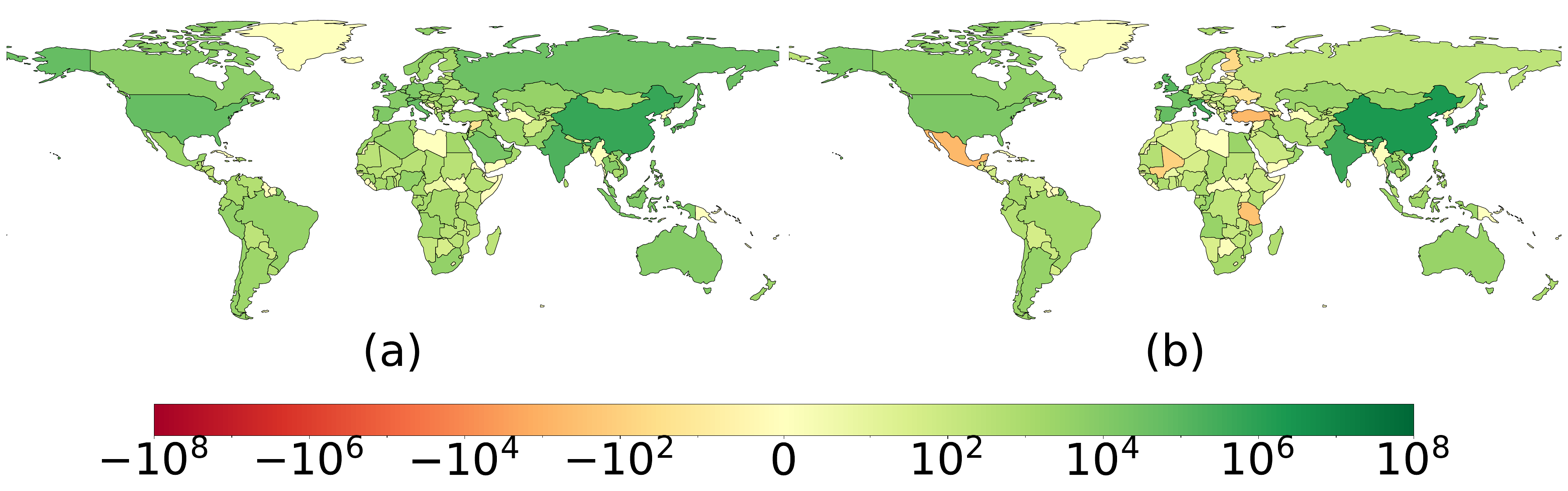}
    
    \caption{Scenario 2 $(\delta=0.3 , \alpha=0.3)$: Heatmap of changes in aggregate (a) final demand and (b) output of countries with respect to Scenario 1 $(\delta=0.3)$. All figures in mUSD. Reallocation parameter $\alpha$ is applied to \texttt{TOP-GAS} outflows.
    }
    \label{fig:heat_alpha}
    
\end{figure}
\begin{figure}
\centering
    \subfloat[]{\includegraphics[width=0.38\linewidth]{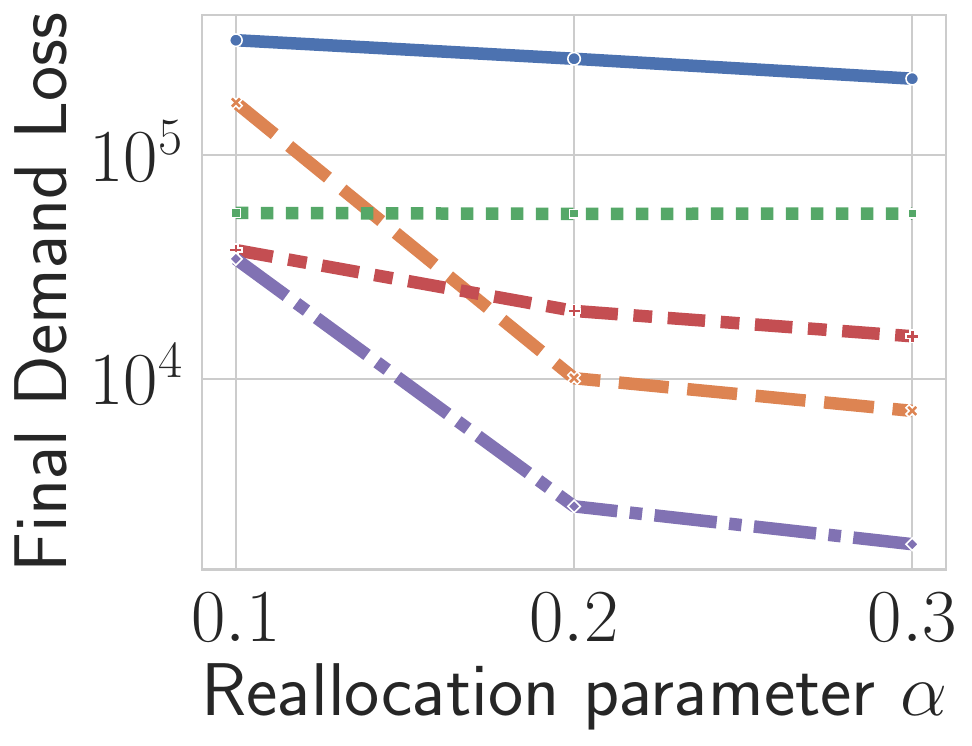}
    \includegraphics[width=0.1\linewidth]{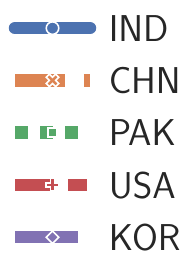}}
    \subfloat[]{\includegraphics[width=0.38\linewidth]{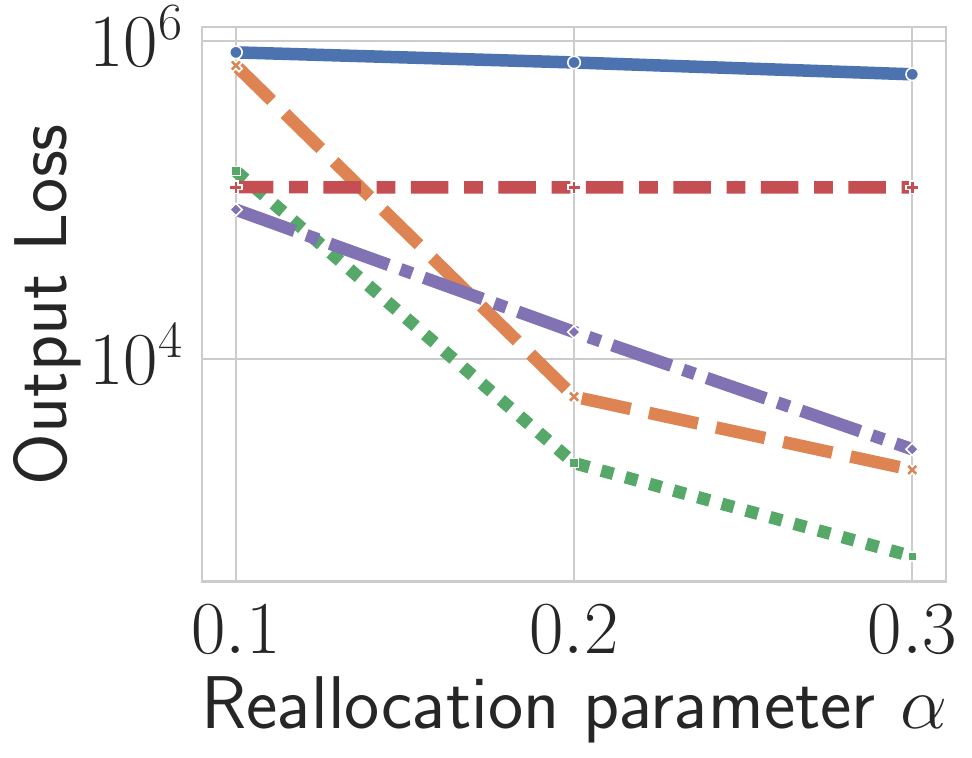}
    \includegraphics[width=0.1\linewidth]{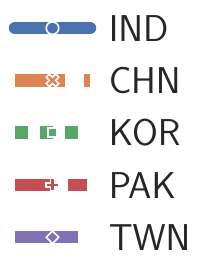}}
    
    \caption{Scenario 2 $(\delta=0.3, \alpha=\text{var.})$: Top 5 most impacted countries (excl. \texttt{GULF}) in terms of loss in (a) final demand, (b) output, with respect to original values with varying reallocation 
    parameter $\alpha$ in \texttt{TOP-GAS} outflows. All figures are in mUSD.}
    \label{fig:vary_alpha}
\end{figure}

\begin{figure}
 \centering   
    \includegraphics[width=1\linewidth]{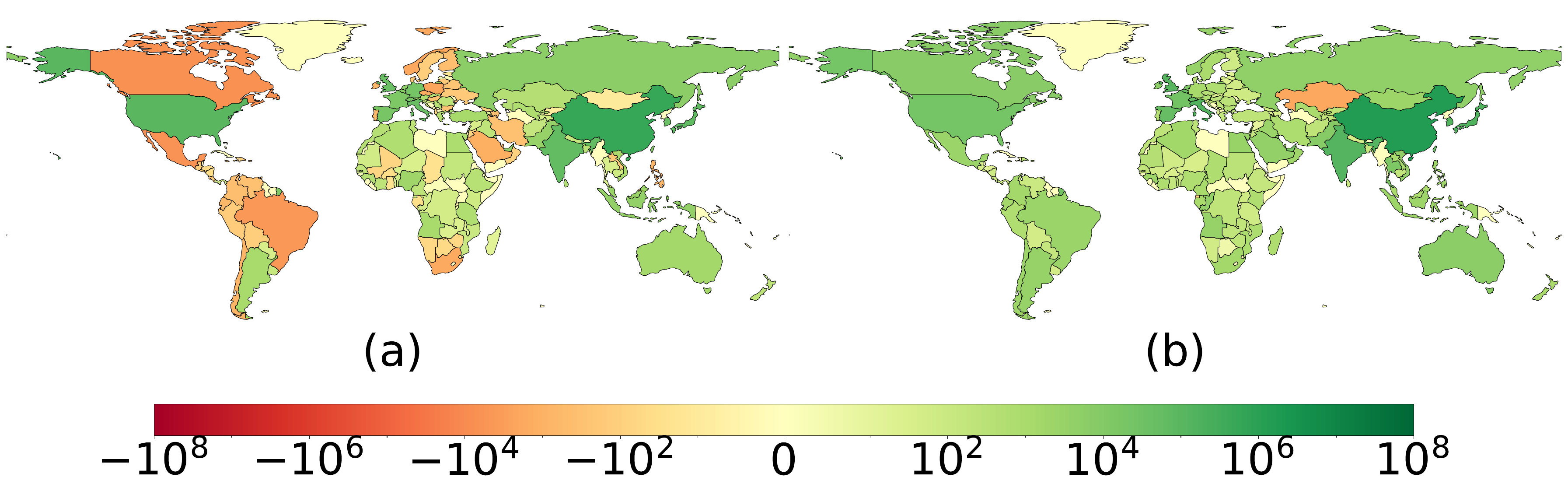}
    
    \caption{Scenario 3 $(\delta=0.3 , \alpha=0.05, \beta=0.05)$: Heatmap of changes in aggregate (a) final demand and (b) output of countries with respect to Scenario 1 $(\delta=0.3)$. All figures in mUSD. Reallocation parameter $\alpha$ is applied to global flows, primary input increase $\beta$ is applied to
    \texttt{TOP-GAS}.}
    \label{fig:heat_beta}
\end{figure}

\begin{figure}
\centering
    \subfloat[]{\includegraphics[width=0.38\linewidth]{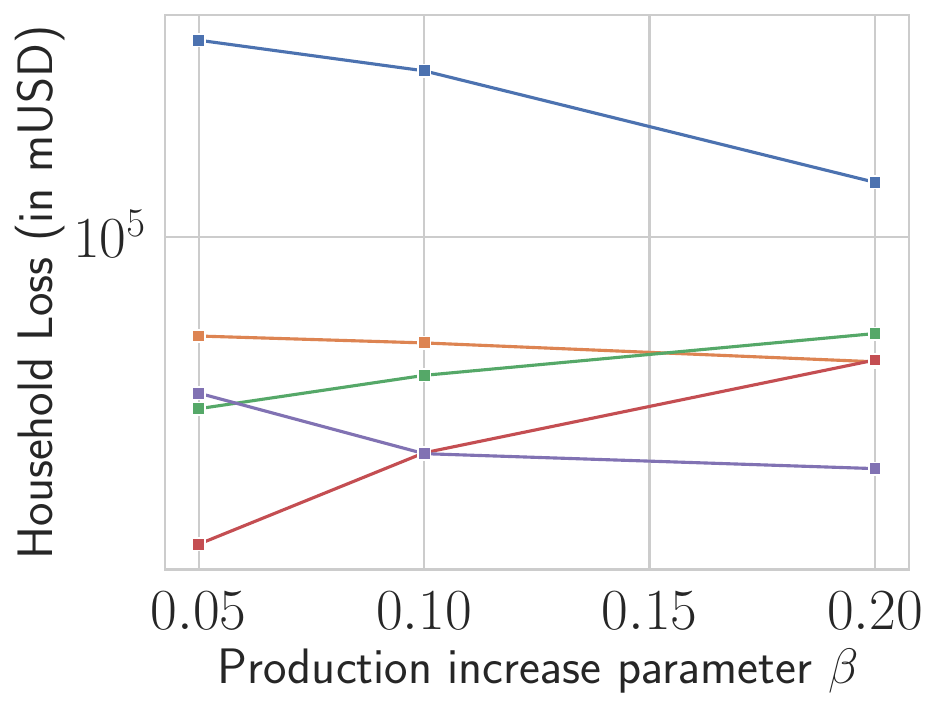}
    \includegraphics[width=0.1\linewidth]{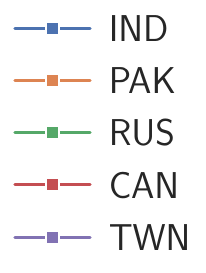}}
    \subfloat[]{\includegraphics[width=0.38\linewidth]{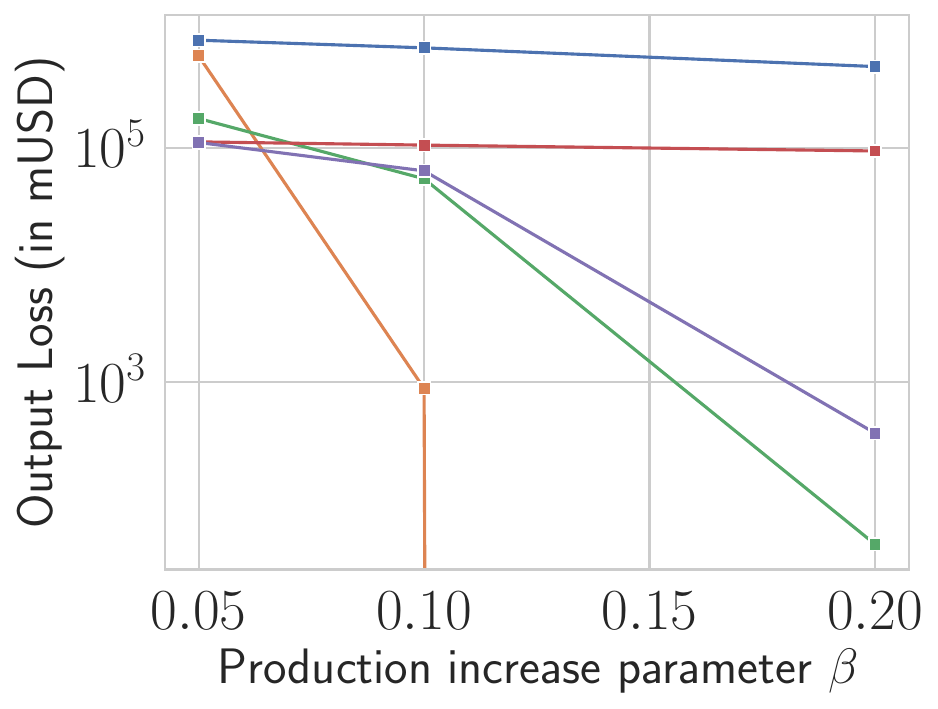}
    \includegraphics[width=0.1\linewidth]{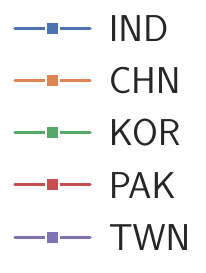}}
    \caption{Scenario 3 $(\delta=0.3, \alpha=\beta,  \beta=\text{var.})$: Top 5 most impacted countries (excl. \texttt{GULF}) in terms of loss in (a) final demand, (b) output, with respect to original values with varying primary input increase 
    parameter $\beta$ in \texttt{TOP-GAS} firms, $\alpha$ applied to all flow capacities. All figures are in mUSD.}
    \label{fig:vary_beta}
\end{figure}
\begin{figure}
\centering
    \subfloat[]{\includegraphics[width=0.48\linewidth]{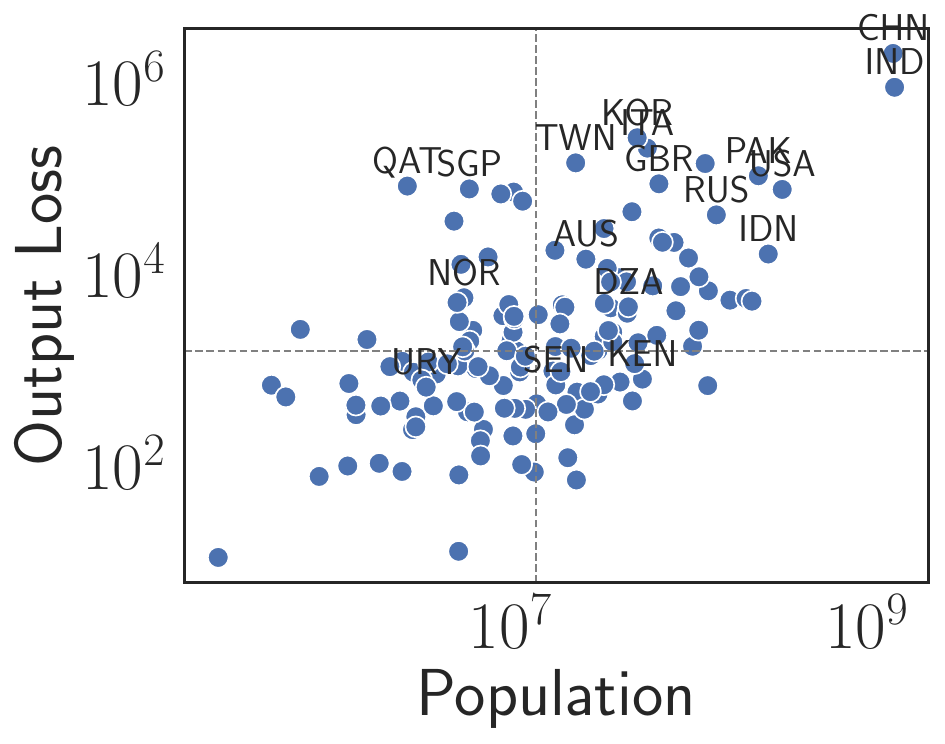}}
    \subfloat[]{\includegraphics[width=0.48\linewidth]{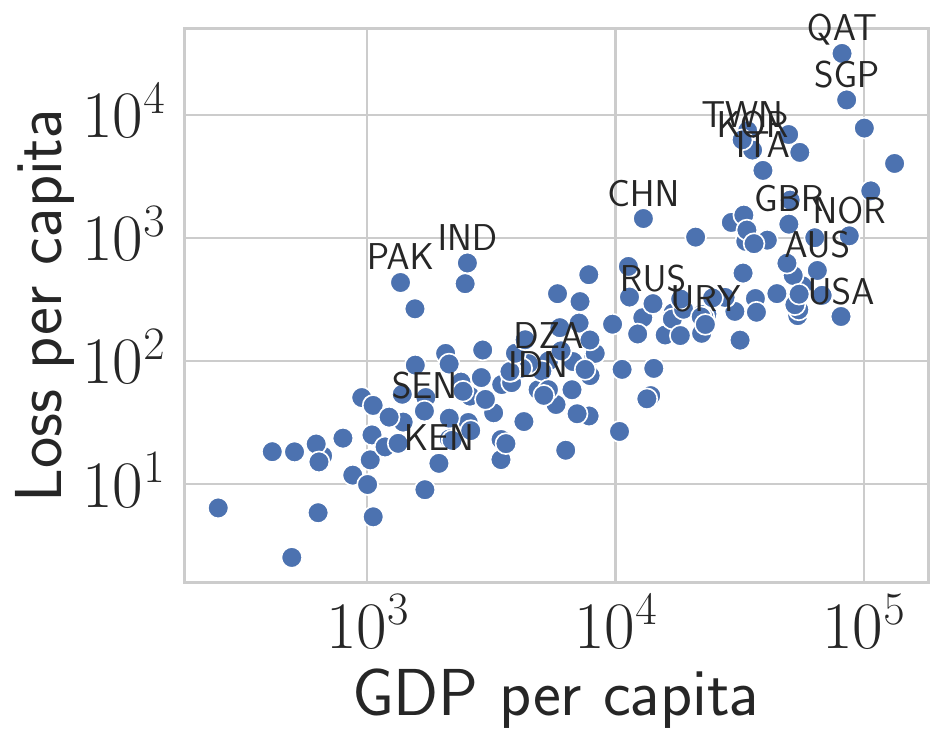}}
    
    \caption{Under Scenario 1 $(\delta=0.3)$:(a)  Output loss with respect to original values (in mUSD) vs population of different countries. The dotted lines are the median population and median loss. (b) Output loss per capita vs GDP (nominal) per capita (in 2023 USD).
    }
    \label{fig:loss_pop}
\end{figure}

\subsection{Disruption propagation networks}

While MRIO models quantify overall impacts, the specific pathways through which disruptions propagate are not immediately captured in the aggregated results.
To better understand how shocks propagate in a highly
interconnected system, it is useful to study disruption propagation networks within the global input–output network. 
We define a \uline{disruption propagation network} as a directed graph consisting of edges $(u, v)$ over which the flow after disruption reduces more than a defined loss threshold.
These networks can help identify critical bottlenecks and amplification nodes where the
shock either intensifies or can be effectively mitigated through
policy interventions.  In Figure~\ref{fig:cascade-graph-hops} (in the appendix), we show the distribution of the shortest path distances for all nodes (i.e., sectors and final demand regions) from the source node, \texttt{QAT-GAS}, in the disruption propagation network. We show that the structure of this network is very different depending on the loss threshold: a lower threshold generates a network which is not only larger and has more nodes, but also more interconnected as evidenced by the narrow shape of the distribution. A lower threshold creates shorter paths from the source to each node in comparison to a higher threshold. For the purpose of our following study, we use a higher loss threshold of 100 mUSD to filter out less important paths of disruption propagation. We study two such partial networks below: one focusing on US final demand and the other on Taiwan's electronic sector.

\begin{figure*}
    \centering
    \subfloat[]{\includegraphics[width=0.48\linewidth]{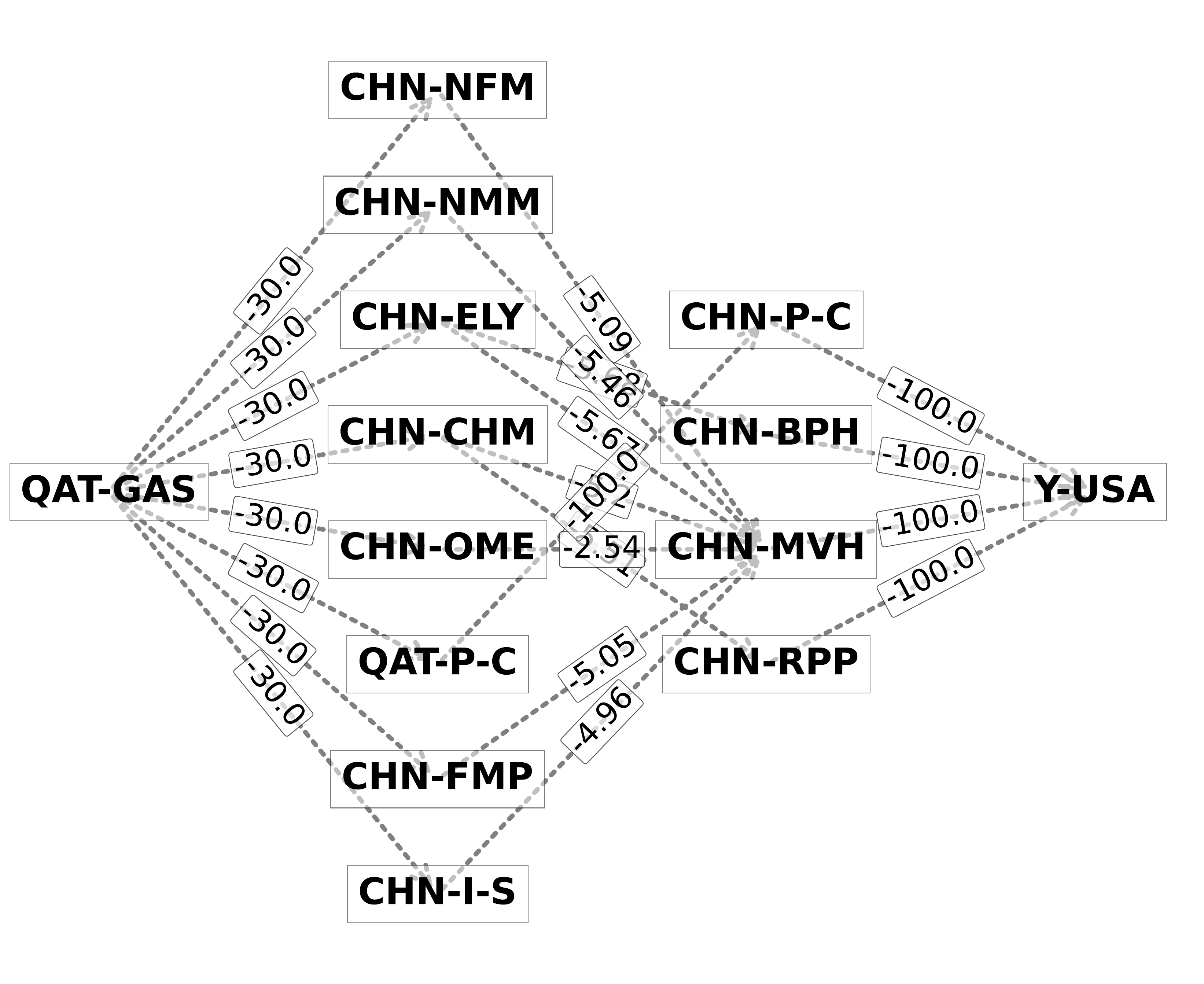}}
    \subfloat[]{\includegraphics[width=0.48\linewidth]{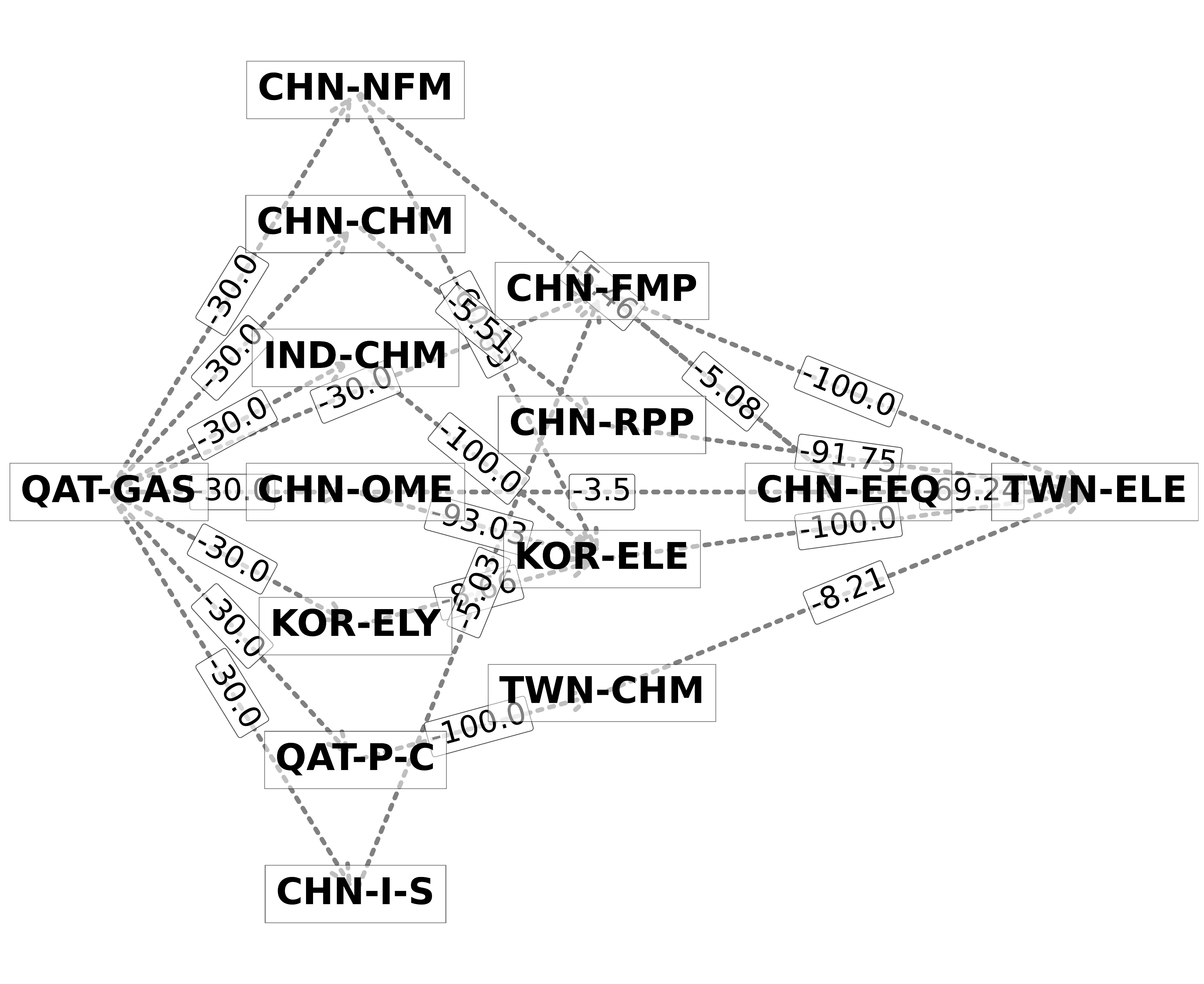}}
    \caption{Top 10 disruption propagation paths by aggregate loss (max length = 3, loss threshold =100) under Scenario 1 $(\delta=0.3)$ for two target nodes: (a) US final demand, (b) Taiwan electronics. Edge-labels show the change in flow as percentages of the original values. 
    }
    \label{fig:cascades}
\end{figure*}

Figure~\ref{fig:cascades} shows the indirect effects of a 30\% disruption in Qatar’s gas sector on chemicals, machinery, electronics, and final demand across economies. The immediate impact falls on China’s chemical sector (CHN-CHM) and Qatar’s petroleum and coal products sector (QAT-P-C). The initial disruption is amplified as it propagates through the network. Despite originating from a 30\% disruption, this leads to a 100\% disruption in the flow from China's Rubber and Plastic Products manufacturing sector (CHN-RPP) to final demand of the US (Y-USA) in the first subgraph, and 91.75\% from CHN-RPP to Taiwan’s electronics sector (TWN-ELE) in the second subgraph. Naphtha, a co-product of LNG production, feeds into petrochemicals and helium into electronic equipment (CHN-EEQ), with losses rippling further to Taiwan’s electronics sector (TWN-ELE). Gas supply constraints also propagate through Taiwan’s electricity sector (TWN-ELY), disrupting electronics manufacturing and affecting Korea’s electronics sector (KOR-ELE). Korea’s electronics sector further suffers losses through China’s machinery sector (CHN-OME) at 93.03\%, reflecting broader network dependence on gas-intensive intermediate production.  These results illustrate that a single supply disruption propagates through multiple channels, including feedstocks, energy supply, and intermediate manufacturing sectors.

Finally, to get a bigger picture of the global trend in losses under scenario 1, we examine correlations of country's population size and output loss; and correlation of loss per capita versus GDP per capita. The results are shown in Figure~\ref{fig:loss_pop}(a) and (b) respectively. The dashed lines in Figures~\ref{fig:loss_pop}(a) indicate the median population and median loss, dividing countries into four groups. Larger, more populous economies tend to experience higher aggregate losses, as seen in the upper-right quadrant where countries like China and India are located. However, the spread of points shows that population alone does not determine impact, some smaller countries also exhibit relatively high losses, reflecting differences in economic structure and dependence on gas-linked supply chains.

Figures~\ref{fig:loss_pop}(b) shows losses on a per capita basis. A positive relationship shows higher-income countries experience larger losses per person. Economies such as Singapore, Qatar, and Taiwan lie toward the upper end, indicating greater exposure per capita, likely due to their deeper integration into global, energy-intensive production networks. Lower-income countries are clustered at lower per capita losses, although some still face meaningful impacts.

These figures highlight the uneven nature of the effects of disruption. Aggregate losses are concentrated in large economies, while per capita impacts are more pronounced in wealthier, highly interconnected countries.

\section{Discussion}
\label{sec:discussion}

Our analysis shows that a disruption to Qatar's gas sector can generate widespread, and sometimes unexpected, cascading effects across sectors and regions through global supply chains. The largest impacts fall on the major Asian economies of India, Pakistan, China, and South Korea, followed by Italy and the United States. The severity of the US impact is notable given that the country has no direct gas-based trade with Qatar and is itself one of the world's largest gas producers, underscoring the reach of indirect transmission channels. Output losses display a similar geographic pattern, with high impact on Taiwan.
We observe that the reallocation of flows leads to significant benefits for final demand, with China, the US, and Korea seeing the largest improvements. 
Finally, production expansion by the top gas producers, and flow reallocation substantially mitigates
losses in some large economies such as China and Korea but even after increasing $\beta$ to 0.2, India and Pakistan's output only partially improve, as shown in Figure~\ref{fig:vary_beta}. This implies that there may be other structural bottlenecks that cannot be overcome by just expanding production by \texttt{TOP-GAS}. 
Our results show that the solution space under increased production and reallocation is very complex, and the final outcomes depend on how countries prioritize output losses versus final demands. For instance, Indian officials gave priority to households over businesses under pressure from gas shortages~\cite{wsj2026indiagas}. In our experiments, we give equal weight to both, i.e., $\lambda_1 = \lambda_2$.
However, our linear programming and disruption propagation network analysis methods provide a framework to systematically explore potential solutions and tradeoffs, by assigning different weights to output and final demand.

Aside from LNG, Ras Laffan jointly produces condensate, helium, liquefied petroleum gas, naphtha, and sulphur, 
which will lead to export reductions of 24\%, 14\%, 13\%, 6\% and 6\% respectively~\cite{CNA2026, Annesley2026}.
These disruptions propagate downstream affecting household energy use and different sectors including 
transportation, petrochemicals, and semiconductors, among others.  
In Japan and South Korea, where naphtha imports are heavily sourced from the Middle East, production of ethylene, 
PVC, and polyethylene, which are inputs to plastics, packaging, and construction materials, have also been 
disrupted~\cite{qatartribune2026}. 
Taiwan and South Korea import approximately 69\% and 64.7\% of their helium, a critical input in semiconductor 
fabrication, from Qatar respectively, making both countries' chip industries directly exposed.
Disruptions in these materials need to be considered along with gas disruptions, in order to understand the full 
economic impact.

This work has several limitations. MRIO captures fixed production relationships and does not account for
price adjustments which means substitution driven by price signals is not
captured. Adaptation is modeled through exogenous increase in flow and
production capacities, and hence 
does not capture real-world frictions like long-term contracts,
strategic reserves, geopolitics etc., and 
assumes reallocation is feasible once capacity is increased. 
Finally, the
linear program provides multiple feasible solutions and does not
guarantee uniqueness.

There are several caveats in the counterfactual scenarios on reallocation and production increase we study here.
Aside from Qatar, the US and Australia are also major LNG producers, but neither can immediately fill the supply 
gap. Both countries are operating close to capacity, with production already committed under long-term 
contracts~\cite{williams2026}. Reallocation is further constrained by energy security priorities. LNG exports are 
not purely market-driven, as governments may intervene to prioritize domestic supply during periods of shortage, 
consistent with policies such as Australia's Domestic Gas Security Mechanism~\cite{ausgovt2026}. 

We also note that there are challenges in studying disruptions in other materials associated with gas production.
For instance, helium supply does not necessarily increase with natural gas production. Helium is only recoverable 
from gas streams with sufficiently high concentrations and requires dedicated processing infrastructure, which 
means that only a limited number of gas fields are commercially viable for extraction~\cite{HuandLi2025}. Despite 
being a major LNG exporter, Australia does not have the infrastructure to capture helium from its gas 
production~\cite{george2026}. These supply-side rigidities mean that disruption cannot be resolved through market 
reallocation alone, and its economic consequences will propagate across global production networks for an extended 
period. LNG liquefaction projects take an average of 4-5 years between the final investment decision and 
completion, with no new capacity expected to come online on a sufficient scale to address the disruption in the 
next 2-3 years~\cite{iea2026lngtracker}.
 While these constraints limit immediate supply-side responses, understanding the potential mitigating effects of 
 partial trade reallocation and capacity expansion remains critical for policy. Thus, we consider scenarios to 
 quantify the global economic impact of disruption and evaluate the extent to which these adaptive responses could 
 reduce its consequences.


\section{Methodology}
\label{sec:method}

\begin{table*}[!ht]
    \centering
    \begin{tabular}{|l|c|p{3.5in}|}
        \hline
            \multirow{7}{2em}{Sets} & $\mathcal{F}$ & Set of all firms \\
            & $S \subseteq \mathcal{F}$ & Set of firms with same product called \textit{sector}\\
            & $\mathcal{F}_{GAS}$ & Set of \texttt{TOP-GAS} firms \\
            & $\mathcal{S}$ & Set of all sectors \\
            & $\mathcal{R}$ & Set of all final demand regions \\
            & $E_I \subseteq \mathcal{F}\times \mathcal{F}$ & Set of all firm-to-firm flow edges\\
            & $E_H \subseteq \mathcal{F}\times \mathcal{R}$ & Set of all firm-to-region flow edges\\
            & $E_{\delta}$ & Set of disrupted edges \\
            & $E_{\alpha}$ & Set of adaptive edges enabling reallocation \\
            \hline
            \multirow{3}{2em}{Variables} & $ x_i~\forall i \in \mathcal{F} $ & Output of each firm $i$ \\
            & $z_{i,j} ~\forall (i,j) \in E_I$& Flow from firm  $i$ to firm $j$\\
            & $h_{i,r} ~\forall (i, r) \in E_H$ & Flow from firm $i$ to final demand region $r$\\
            & $\epsilon_{S,i} ~\forall (S, i) \in \mathcal{S}\times \mathcal{F}$ & Wasted input from sector $S$ to firm $i$ \\
            \hline
            \multirow{7}{*}{Constants}& $A_{S,i} ~\forall S\in \mathcal{S}, i \in \mathcal{F}$ & Input coefficient of firm $i$ from sector $S$\\
            & $B_i ~\forall i \in \mathcal{F}$ & Primary input coefficient of firm $i$ \\
            & $P_i^*, Z_{i,j}^*, H_{i,r}^*, X_i^*$ & {MRIO baseline values for primary inputs, intermediate flows, final demand flows, firm outputs respectively}\\
            & $Z^{\delta}_{i,j}, H^{\delta}_{i,r}$ & Scenario 1 values for intermediate flows and final demand flows respectively.\\
             & $\delta$ & Flow capacity disruption parameter \\
             & $\alpha$ & Flow capacity reallocation parameter \\
            & $\beta$ & Primary input capacity increase parameter\\
            & $W_{max}$ & Max allowable wasted flow for a (sector, flow) pair\\

            \hline
    \end{tabular}
    \caption{List of symbols used in the linear program described in Table~\ref{tab:lp}.}
    \label{tab:lp_symbols}
\end{table*}
\begin{table*}[!ht]
    \centering
    \begin{tabular}{l l r}
         maximize & \multicolumn{2}{l}{ $\underbrace{\displaystyle \sum_{i\in \mathcal{F}} x_i}_{\text{Production incentive}}  +  \underbrace{\lambda_1 \displaystyle \sum_{(j,k)\in E_I} z_{j,k}}_{\text{Intermediate flow incentive}} +  \underbrace{\lambda_2 \displaystyle\sum_{(l,r)\in E_H} h_{l,r}}_{\text{Final demand flow incentive}}
         -\underbrace{\lambda_3\displaystyle\sum_{(S,m) \in \mathcal{S}\times \mathcal{F}} \epsilon_{S,m}}_{\text{Penalty for wasted inputs}}$}  \\
         subject to, & & \\
         \multirow[t]{2}{*}{(1) Leontief constraints:} & $A_{S,i}x_i + \epsilon_{S, i} =  \displaystyle \sum_{j\in S; (j,i)\in E_I} z_{j,i}$  & $\forall S\in \mathcal{S}, i\in \mathcal{F} $\\ 
         & $B_i x_i \leq P_i^*$ & $\forall i\in \mathcal{F}\setminus \mathcal{F}_{GAS}$\\ 
         & $B_i x_i \leq (1+\beta)P_i^*$ & $\forall i\in \mathcal{F}_{GAS}$\\
         (2) Conservation of flow: & $\displaystyle\sum_{(i,j) \in E_I} z_{i,j} + \displaystyle\sum_{(i,r)\in E_H} h_{i,r} = x_i$ & $\forall i \in \mathcal{F}$ \\ 
         (3) Flow capacity bounds: & $z_{i,j} \leq Z^*_{i,j}$ & $\forall (i,j) \in E_I\setminus (E_\delta \cup E_\alpha)$\\
         & $h_{i,j} \leq H^*_{i,j}$& $\forall (i,j) \in E_H\setminus (E_\delta \cup E_\alpha)$ \\
         & $\epsilon_{S,i} \leq W_{max}$\\
         \multirow[t]{4}{*}{(4) Disruption \& adaptation:} & $z_{i,j} \leq (1-\delta)Z_{i,j}^*$ & $\forall (i,j) \in E_I\cap E_{\delta}$ \\
         & $h_{i,r} \leq (1-\delta)H_{i,r}^*$ & $\forall (i,r) \in E_H\cap E_{\delta}$ \\
         & $(1-\alpha) Z_{i,j}^\delta \leq z_{i,j}\leq (1+\alpha)Z_{i,j}^*$ & $(i,j) \in E_I \cap E_{\alpha}$  \\
         &  $ (1-\alpha) H^{\delta}_{i,r} \leq h_{i,r}\leq (1+\alpha)H_{i,r}^*$ & $(i,r) \in E_H \cap E_{\alpha}$
        \\
    \end{tabular}
    \caption{The linear program for the study of trade disruptions and adaptations. The symbols are described in Table~{\ref{tab:lp_symbols}}.}
    \label{tab:lp}
\end{table*}

\subsection{Input-Output Network}
We employ a multiregional input-output (MRIO) model~\cite{millerBlair2009}. The MRIO table is obtained from the newly released GTAP 12 database for the year 2023, which covers 163 countries and 65 sectors~\cite{aguiar2025global}. We consider an economy consisting of production sectors, denoted by $\mathcal{S}$, and final demand regions or countries, denoted by $\mathcal{R}$. Each sector $S\in \mathcal{S}$ produces a unique good, e.g., steel, gas, etc. A \textit{firm} is a sector in a country, identified by a tuple (country, sector), e.g. USA-GAS for US gas industry, CHN-CHM for China chemicals industry. The set of all firms is denoted by $\mathcal{F}$. Note that each firm belongs to a global sector along with other firms producing the same good, e.g., USA-GAS, RUS-GAS, CAN-GAS etc. all belong to the global GAS sector.
A firm produces its goods using intermediate inputs from other firms and primary inputs such as labor, capital etc. The gross output $x_i$ of a firm $i$ is the sum of its intermediate inputs and value added.

We model the flow of goods between firms and end-users as a directed capacitated network. In this network, each firm is a node  and each country has one representative node for its final demand.  There are two types of edges: firm-to-firm edges forming the set $E_I$, and firm-to-final demand edges forming the set $E_H$. Each edge carries a flow of goods whose maximum value (in USD) cannot exceed the stated edge capacity (in USD).
Each firm sends its goods along this network to other firms (intermediate flows) and final demand regions (final demand flows). The flow value from firm $i$ to $j$ is denoted by $z_{i,j}$; flow from firm $i$ to final demand region $r$ is denoted by $h_{i,r}$. In the Leontief framework, the core assumption is that each firm operates on fixed production recipes based on known input coefficients. In our model, products of the same sector sourced from different countries are substitutable. Thus, an input coefficient $A_{S,i}$ is computed as the total amount of input required from the firms in sector $S$ to produce one unit of output in firm $i$, 
\begin{equation}A_{S,i}= \sum_{j\in S; (j,i)\in E_I}z_{j,i}/x_i \label{eqn:ln_func}\end{equation} All of the output of a firm is distributed to other firms and final demand as, 
\begin{equation}x_i = \sum_{(i,j) \in E_I} z_{i,j} + \sum_{(i,r) \in E_H} h_{i,r}\label{eqn:dist}\end{equation}.

\subsection{LP formulation}
We model a disruption or an adaptation as a change in flow-capacities for a given subset of edges. Given a disruption, our goal is to find an equilibrium, i.e., the allocation of flows ~$\{z_{i,j}: (i,j) \in E_I\}, \{h_{i,r}: (i,r) \in E_H\}$ and firm outputs~$\{x_i: i\in \mathcal{F}\}$ on the IO network such that the solution is optimal according to a given objective. We adopt an optimization approach where we want to find the flow allocations and firm outputs which maximize the total output and total final demand flows in the network. We formulate this bi-objective problem as a weighted combination with parameters $\lambda_1$ and $\lambda_2$ controlling the importance given to intermediate demand flows versus final demand flows. In this study we maintain $\lambda_1 = \lambda_2$, i.e., give equal weight to both intermediate and final demand.
Due to the linearity of the Leontief framework, this problem can be formulated as a linear program as shown in Table~\ref{tab:lp}. (1) The Leontief constraints ensure that the solution flows adhere to the production function defined for every pair of sectoral input flow and firm production (as shown in Equation~\ref{eqn:ln_func}). (2) The conservation of flow constraints ensure that everything that is produced by any firm is distributed to other firms and final demand regions along its outgoing edges, i.e., there is no wasted production (as shown in Equation~\ref{eqn:dist}). (3) We have the flow capacity bound  for each edge $(i,j)$ which is set to the original MRIO value, $Z^*_{i,j}$, by default. (4) However, if it is in the disrupted subset of edges, the capacity is reduced by $\delta$ fraction of the original value. 
And if it belongs to the adaptive edges (Scenario 2 \& 3), then the flow is constrained to be in the interval $[(1-\alpha)Z^{\delta}_{i,j}, (1+\alpha)Z^*_{i,j}]$.
Finally, for scenario 3, we also allow increase in primary input for selected firms by $\beta$ fraction.

We find that the above LP is difficult to solve using the Gurobi solver. 
We introduce an approximation to speed up the computation by aiding the LP solver to reach convergence faster.
We modify the standard Leontief equilibrium condition to allow slack in the production, i.e., sectoral input flows need not be fully utilized by the current production level. 
For each sector $S$ and firm $i$ pair, this is given by~$\epsilon_{S,i} = \sum_{j\in S} z_{j,i} - A_{s,i}x_i$. 
Each wasted input flow is limited to a  maximum value, $W_{max}$, and is penalized in the objective to move the solution towards minimizing overall wasted flows. We observe a tradeoff in speed and accuracy when we increase $W_{max}$ from 0 to 100 mUSD, the computation time reduces by up to 20 times~\ref{tab:waste_eps}, while the total wasted input flow increases to 0.08\% of the total flows $\sum_{(i,j)\in E_I}z_{i,j}$. For the penalty term in the objective, we control the scale of the penalty using a parameter $\lambda_3$, set to a small value $10^{-6}$, so that it doesn't distort the solution space while discouraging wasted flows. 

\subsection{Experimental setup}
\paragraph{Dataset}
The multiregional input-output (MRIO) table is obtained from the newly released GTAP 12 database, which covers 163 countries and 65 sectors with 2023 as the reference year~\cite{aguiar2025global}. We construct the table from the database using the same tools as used in prior works~\cite{guanGlobalSupplychainEffects2020}. Each country has a representative firm for each sector, and a representative household for final demand. The number of firms in our model is 10,595. We treat the flows in this table (also referred to as the original values) as the edge capacities, except where it has been disrupted or is adaptive in our scenarios.

\paragraph{Scaling to large networks}
Given that the MRIO table has  flows ranging from $10^{-12}$ to $10^{6}$ mUSD, we construct the IO network using some thresholds. The total number of possible flows could be in the order of hundreds of millions, which leads to an explosion in the number of flow variables in the LP model. To improve the tractability of the optimization problem, we introduce a flow threshold: a flow between any two nodes with an MRIO value smaller than $\tau_{flow} = 0.01$ mUSD is ignored. This reduces the number of edges (and thus, flow variables) while preserving the significant links in the network. Finally, the input coefficients $\{A_{S,i}\}$ also have a wide range from $10^{-8}$ to $10^{5}$. We introduce a threshold that ignores any input coefficients smaller than $\tau_{ic} = 10^{-5}$, reducing the number of Leontief constraints. 
From a modeling perspective, very small value of a flow or input coefficient is treated as a signal that the input is easily substitutable or otherwise not critical to production in a strict Leontief production limiting sense, justifying the use of these thresholds.

\paragraph{Evaluation on scenarios }
We evaluate our methods on the three scenarios under the defined disruption and analyze the allocation of flows and outputs for the firms in the IO network. For this, we select a range of values for each parameter $\delta, \alpha, \beta$ to evaluate each scenario as given in Table~\ref{tab:scenario}. To model the disruption coming from \texttt{QAT-GAS}, we set the $\delta$ parameter in the LP in Table~\ref{tab:lp} to reduce the upper bound, i.e, capacity, for the subset of edges which originate at \texttt{QAT-GAS} and terminate at any other node, including final demand nodes. In Scenario 1, we set only $\delta$ to a non-zero value, while $\alpha = \beta = 0$, i.e., no edge is made adaptive. In Scenario 2, in addition to a non-zero $\delta$, we set $\alpha$ to allow reallocation of flows over the subset of edges which originate from any firm in \texttt{TOP-GAS} group and terminate at any other node, including final demand regions. We set $\beta = 0$. In Scenario 2, we set all three parameters to non-zero values where $\alpha$ is set for all edges in the network, and $\beta$ is set only for the \texttt{TOP-GAS} firms. 

We analyze the LP solutions first by studying the gas flows. We collect all the flows emerging from  firms in the GAS sector. For each country, we aggregate these gas flows from a GAS firm that is being consumed by any firm within the country or by its representative final demand node. This is called the total gas flow value from a GAS firm being consumed by that country. We further aggregate these country-wise gas inflows into broader regional gas inflows (e.g., US, CAN, MEX etc. become North America)  and plot these gas flows as heatmaps in Figure 1. To analyze the broader impact of the gas flow distribution on countries, we aggregate the  outputs $\{x_i\}$ of all firms in a given country to obtain the country gross output. Next, we aggregate all the flows into the final demand node for each country to obtain the aggregate final demand. Finally, to plot the changes between scenarios, we take the difference between these aggregate values for each country and plot them as heatmaps.

To construct the disruption propagation network, we first select a threshold for the flow loss above which we include an edge as part of the disruption network. Given this network, we extract partial subgraphs for a given target node for analysis using the following method: Find all the paths, i.e, sequence of edges, in the disruption network that begin at the source node, i.e, \texttt{QAT-GAS}, and terminate at the target node under analysis, e.g., US final demand node. We limit the length of such paths to 3 edges in our experiments both for computational reasons, and for analyzing the shorter more direct paths. We compute the weight of a path by summing the losses over each constituent edge. We sort these paths by weight in descending order, and report the top 10 paths in our results.

All hyperparameters settings used in our experiments are given in Table~\ref{tab:hyp_param} (moved to the appendix).

\section*{Acknowledgments}
This work is partially supported by University of Virginia Strategic Investment
Fund Awards: SIF160, SIF186, the Contagion Science P2PE fund SIF176A, NSF grants
CCF-1918656 and CNS-2317193. The  research is also partly supported by the National Security Data \& Policy Institute, Contracting Activity 2024-24070100001.

\bibliographystyle{plain}
\bibliography{references}
\clearpage
\appendix
\onecolumn
\begin{center}
    \fbox{{\textbf{\Large{Technical Appendix}}}}
\end{center}

\paragraph{Additional results}
Figure~\ref{fig:gas_flow_mat_htmp_det} shows the gas flow from producer firms to more fine-grained consumer regions. For clarity, we combined some of the regions into a single region based on low prominence pertaining to gas flows, in the previous plot in main text (Figure~\ref{fig:gas_flow_mat_htmp}). We visualize these flows as Sankey diagrams in Figure~\ref{fig:sankey_gas}.

In Figure~\ref{fig:casc_loss_thresh}, we show how the size of the disruption propagation network, i.e., the number of nodes included, varies with increasing loss threshold, $\tau_{loss}$. This shows that with a low enough $\tau_{loss}$, a large majority ($> 75\%$) of the network participates in the disruption propagation. Yet, this number quickly decreases as $\tau_{loss}$ increases, ultimately retaining only those nodes which propagate major flow losses. In Figure~\ref{fig:casc_kor}, we show the top 10 propagation paths for the firm, Korea electronics.

\paragraph{Additional details on methods}
We report the hyperparameter settings used in our methods in Table~\ref{tab:hyp_param}. For the input coefficient threshold used in our network construction, we report the tradeoff between accuracy in the resulting solutions and time taken to solve in Table~\ref{tab:tau_ic}. Next, we study various settings of $W_{max}$ and $\lambda_3$ and report the accuracy vs. time to solve for the solutions in Table~\ref{tab:waste_eps}.


\begin{table*}[!h]
    \centering
    \begin{tabular}{|c|p{3in}|c|}
    \hline
        Symbol & Description & Value \\
        \hline
         $\tau_{flow}$& Flow Threshold & $0.01$ mUSD\\
         $\tau_{ic}$& Input Coefficient Threshold & $10^{-5}$ \\
         
         $\lambda_1$ & Weight for total intermediate demand flows in LP objective & $1$ \\
         $\lambda_2$ & Weight for total final demand flows in LP objective & $1$ \\
         $\lambda_3$ & Weight for penalty term for wasted input flows in LP objective & $10^{-6}$ \\
         $W_{max}$ & Upper bound for an individual input across sector-to-firm pair & $100$ mUSD \\
         $\tau_{loss}$ & Loss Threshold (for the disruption propagation network) & $100$ mUSD \\
         - & Max. Disruption Path Length & 3 edges \\
         \hline
    \end{tabular}
    \caption{Hyperparameter settings for our model used in the Scenarios}
    \label{tab:hyp_param}
\end{table*}

\begin{table*}[!h]
    \centering
    \begin{tabular}{|c|c|c|c|}
    \hline
         Threshold & Aggregate Error \% &   Mean Error \% (MAPE) & Median Error \% \\
         \hline
         0.001 & 0.596 & 0.508 & 0.028 \\
         $10^{-5}$ & 0.003 & 0.002 & 0.0001 \\
         $10^{-7}$ & $5.85 \times 10^{-6}$ &  $2.98 \times 10^{-6}$ & 0.0 \\
         \hline
    \end{tabular}
    \caption{Scenario 1 $(\delta=0)$: Error in country outputs using varying levels of input coefficient threshold, $\tau_{ic}$, in our method.}
    \label{tab:tau_ic}
\end{table*}

\begin{table*}[htb]
    \centering
    \begin{tabular}{|c |c |c | c | c|}
    \hline
         $\epsilon_{s,i} \leq  W_{max}$ & Penalty Scale ($\lambda_3$) & Total Production & Wasted Input Flows $(\sum \epsilon_{s,i})$  & Time to solve\\
         \hline
         0 & No slack allowed & 187,105,695 & 0 (0.00\%) & 8 hrs \\
         N.B.& 0 &   187,792,119 & 2883500 (1.54\%) & 10 min  \\
         N.B.& $10^{-6}$ & 187,792,119 & 635486.47 (0.34\%) & 10 min  \\
         N.B.& 1 &   187,792,100 & 629593.56 (0.34\%) & 20 min \\
         100 & $10^{-6}$ & 187,250,102 & 147,036 (0.08\%) & 20 min \\
         10 & $10^{-6}$ & 187151721 & 43416 (0.02\%) & 1.50 hr \\
         \hline
    \end{tabular}
    \caption{Tradeoff between speed and accuracy using varying penalty term parameters, $(W_{max}, \lambda_3)$. N.B. stands for no bound. All figures are in mUSD.}
    \label{tab:waste_eps}
\end{table*}

\begin{figure*}[h]
    \centering
    \includegraphics[width=1.0\linewidth]{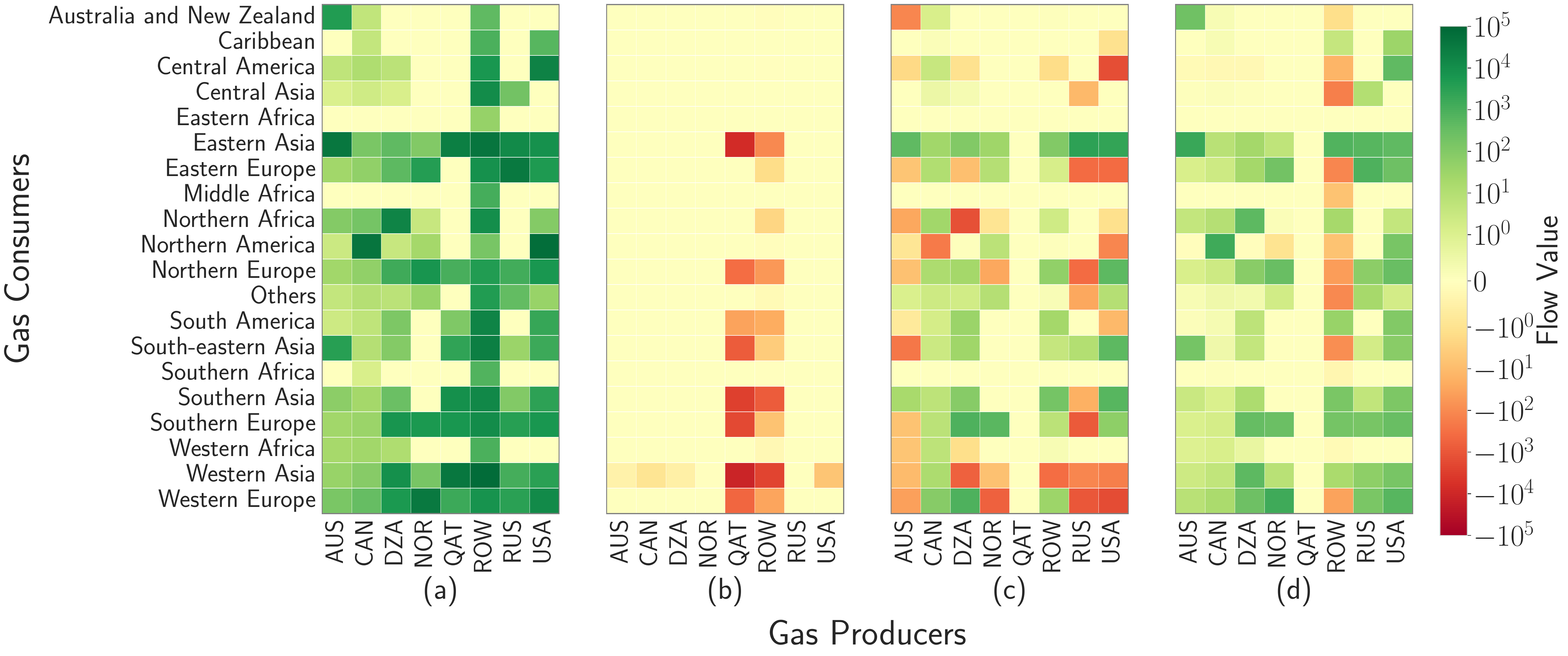}
    \caption{Heatmap of gas flows from producers to consumer regions (more fine-grained groupings) (a) Original flows $(\delta=0)$. (b) Decline in flows at $(\delta=0.3)$ w.r.t original values $(\delta=0)$. (c) Change in flows in Scenario 2 $(\delta=0.3, \alpha=0.3)$ w.r.t. Scenario 1 $(\delta=0.3)$. (d) Change in flows at Scenario 3 $(\delta=0.3, \alpha=0.05, \beta=0.05)$ w.r.t Scenario 1 $(\delta=0.3)$. All figures are in mUSD.}
    \label{fig:gas_flow_mat_htmp_det}
\end{figure*}

\begin{figure*}[h]
    \centering
    \includegraphics[width=1.0\linewidth]{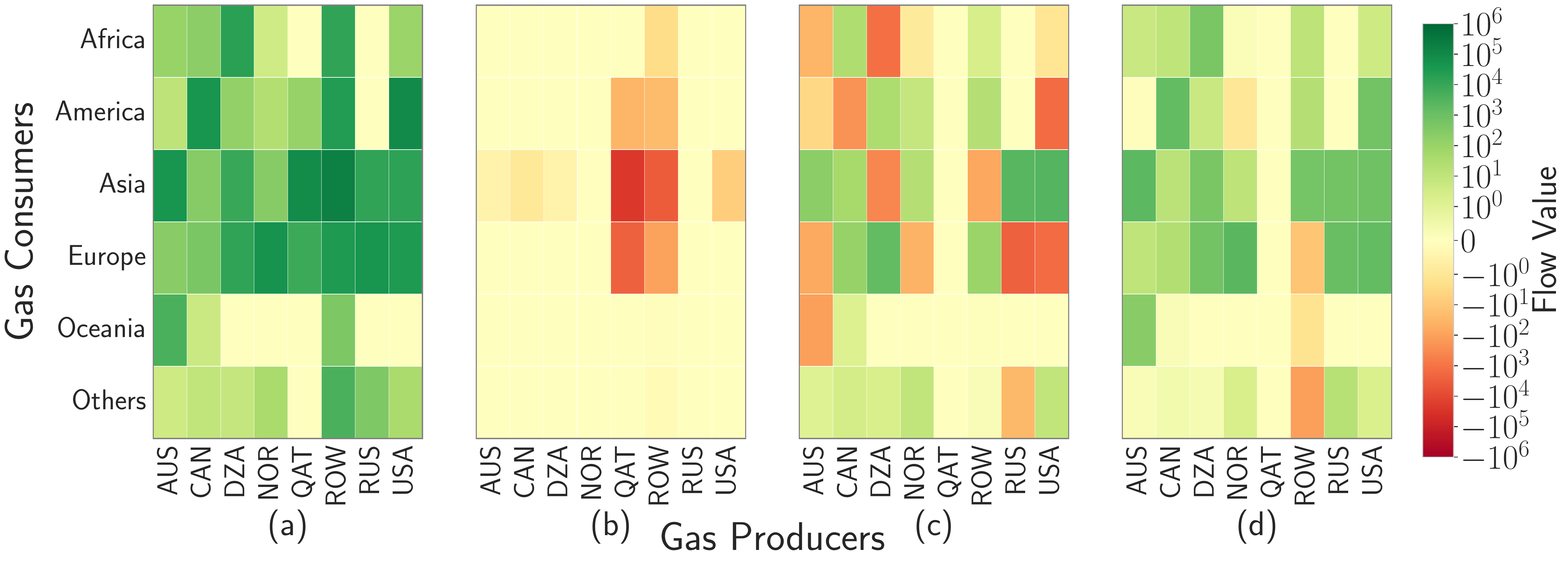}
    \caption{Heatmap of gas flows from producers to consumer continents: (a) Original flows $(\delta=0)$. (b) Decline in flows at $(\delta=0.3)$ w.r.t original values $(\delta=0)$. (c) Change in flows in Scenario 2 $(\delta=0.3, \alpha=0.3)$ w.r.t. Scenario 1 $(\delta=0.3)$. (d) Change in flows at Scenario 3 $(\delta=0.3, \alpha=0.05, \beta=0.05)$ w.r.t Scenario 1 $(\delta=0.3)$. All figures are in mUSD.}
    \label{fig:gas_flow_mat_htmp_cont}
\end{figure*}
    


\begin{figure*}[h]
    \centering
    \subfloat[]{\includegraphics[width=0.48\linewidth]{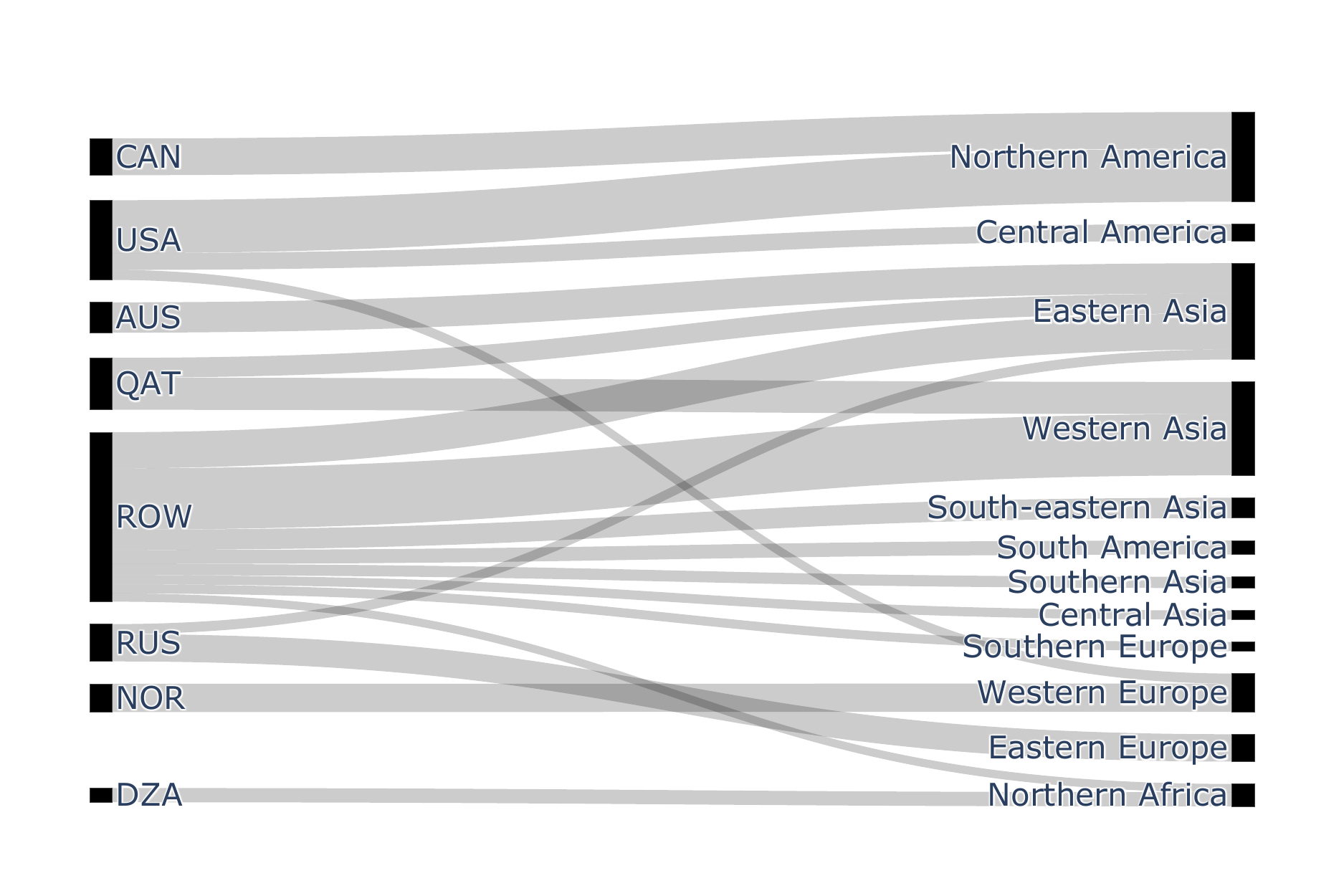}}
    \subfloat[]{\includegraphics[width=0.48\linewidth]{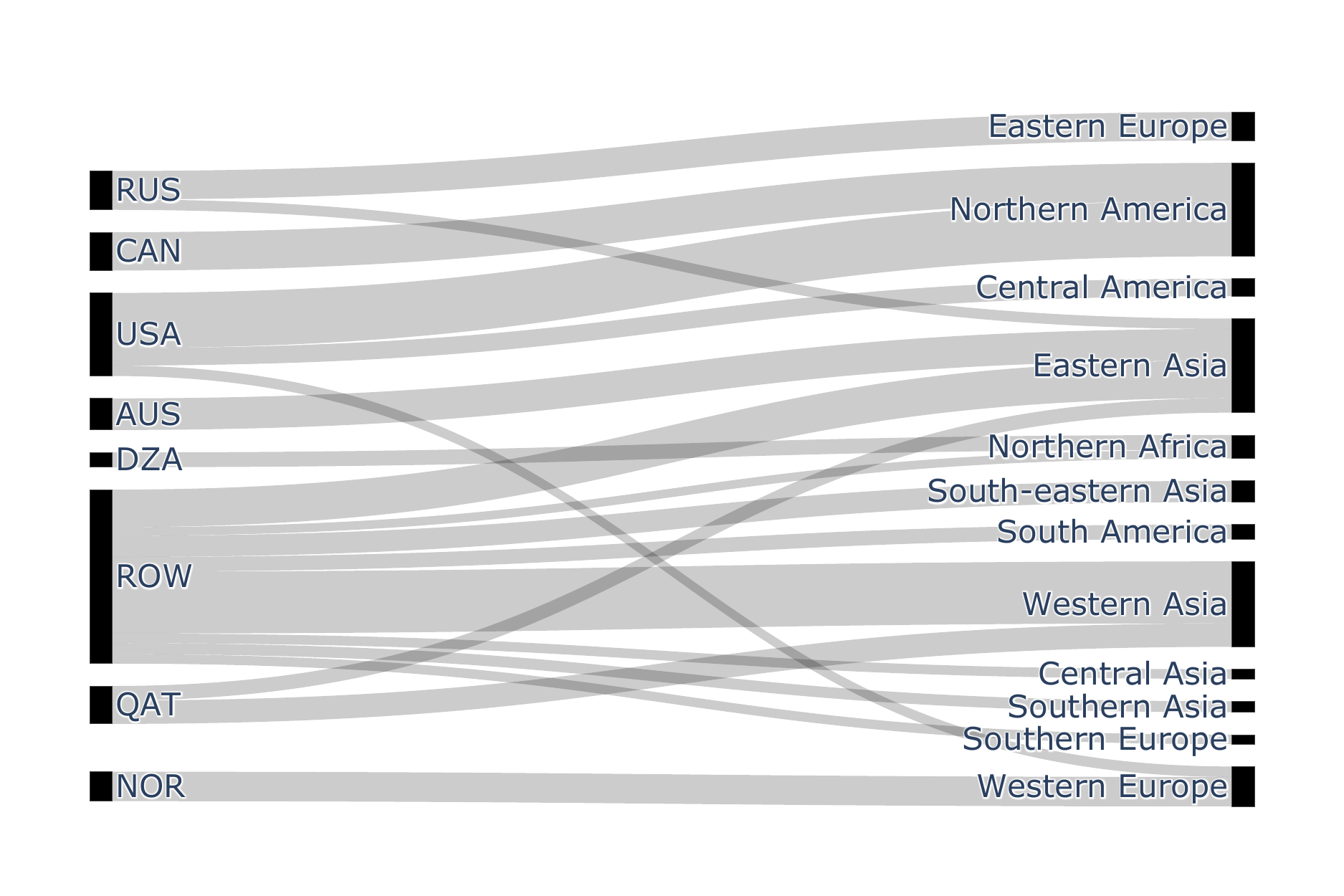}}
    \caption{Gas Flows from Producers to Consumer Regions (a) Original values $(\delta=0)$, (b) Scenario 1 $(\delta=0.3)$. For clarity, only gas flows valued above 100 mUSD are drawn here.}
    
    \label{fig:sankey_gas}
\end{figure*}

\begin{figure*}[!htb]
    \centering
    \includegraphics[width=0.3\linewidth]{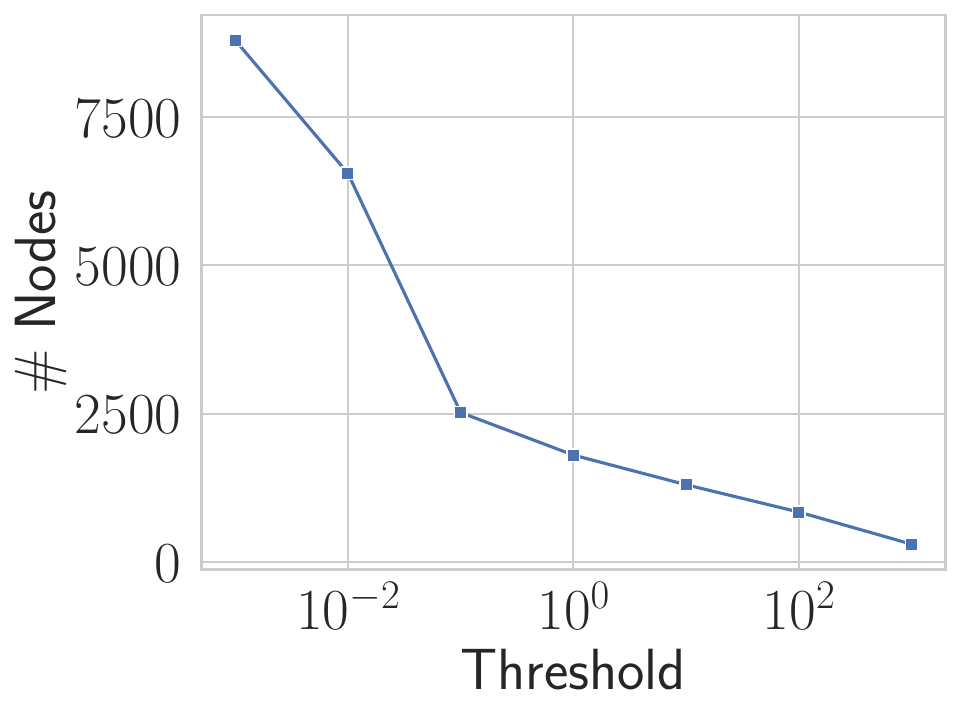}
    \caption{Scenario 1 $(\delta=0.3)$: Size of the disruption propagation network with varying loss threshold $\tau_{loss}$ (in mUSD).}
    \label{fig:casc_loss_thresh}
\end{figure*}

\begin{figure*}[!htb]
    \centering
    \includegraphics[width=0.28\linewidth]{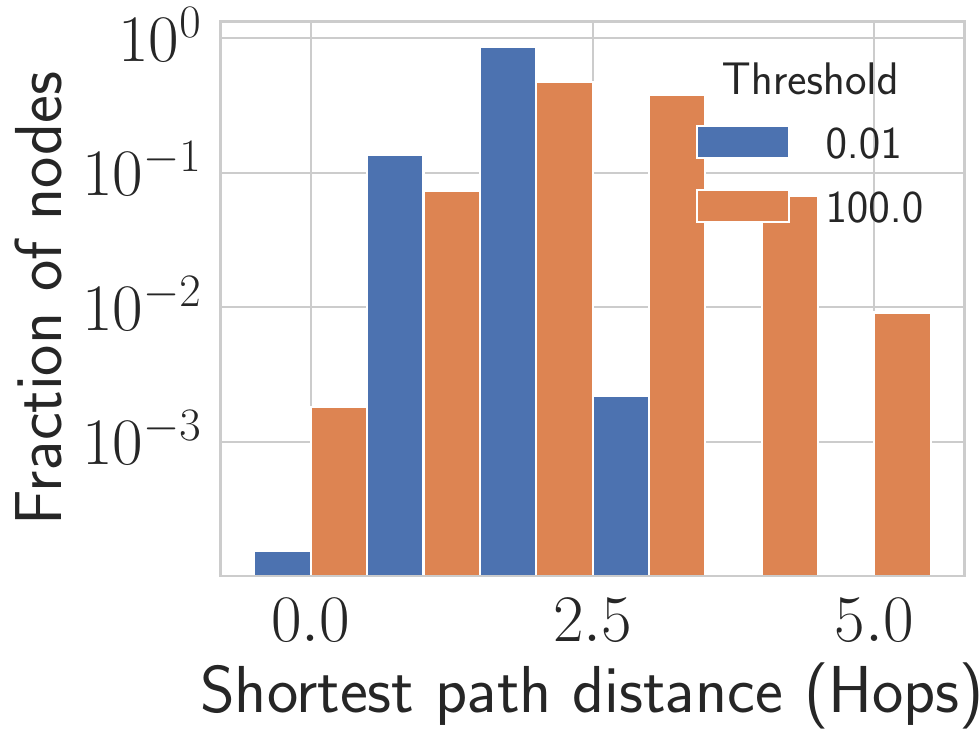}
    \caption{Distribution of shortest path distances from \texttt{QAT-GAS} to nodes in the disruption propagation network for two loss thresholds, $\tau_{loss}= 0.01$ mUSD (blue bars), $\tau_{loss} = 100$ mUSD (orange bars).}
    \label{fig:cascade-graph-hops}
\end{figure*}

\begin{figure*}[!h]
    \centering
    \includegraphics[width=0.5\linewidth]{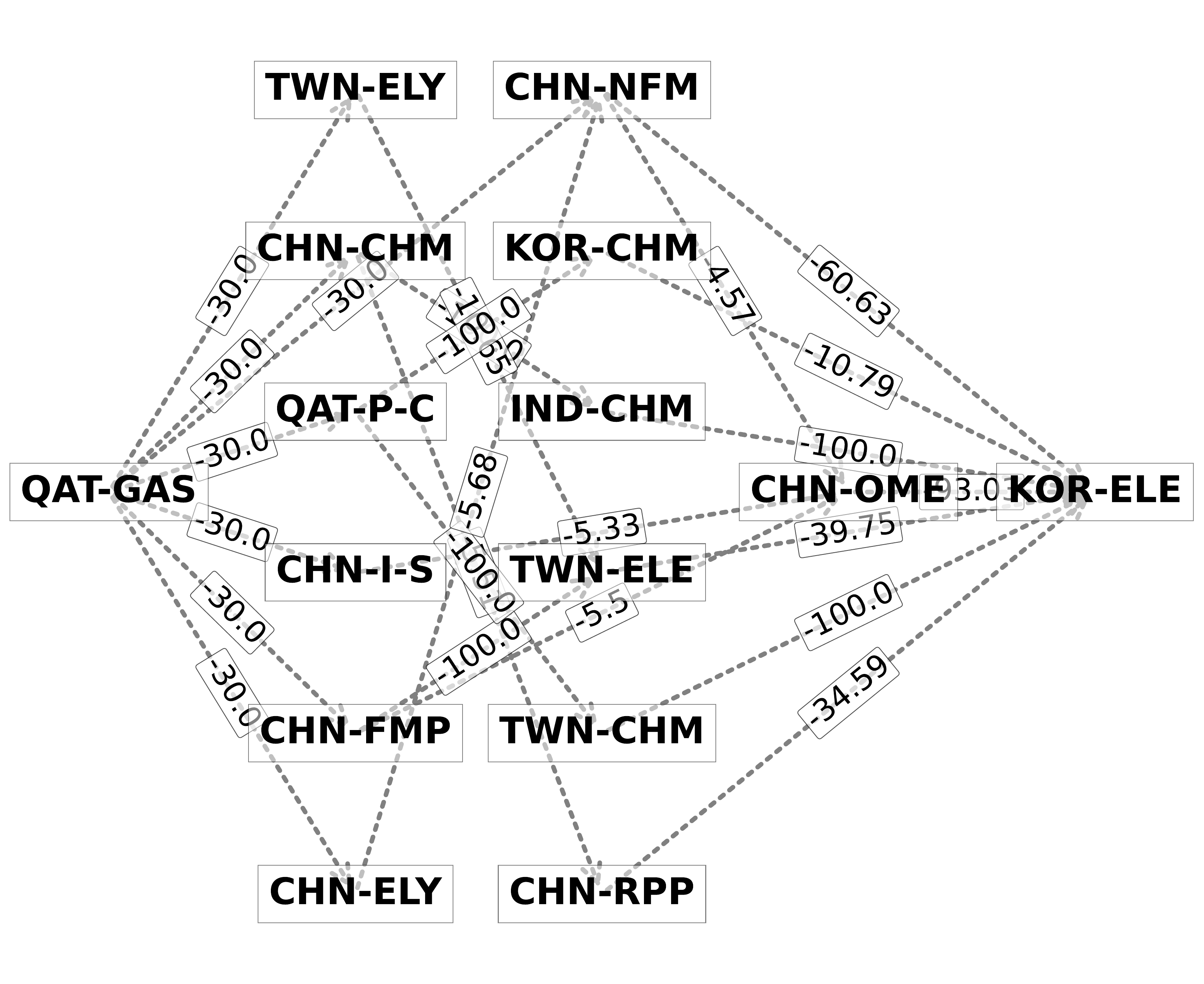}
    \caption{Top 10 disruption propagation paths by aggregate loss under Scenario 1 ($\delta=0.3$) for Korea electronics. The edge-labels show percentage decline in flows from original values. }
    \label{fig:casc_kor}
\end{figure*}

\end{document}